\documentclass[fleqn,usenatbib]{mnras}

\usepackage{newtxtext,newtxmath}

\usepackage[T1]{fontenc}

\DeclareRobustCommand{\VAN}[3]{#2}
\let\VANthebibliography\thebibliography
\def\thebibliography{\DeclareRobustCommand{\VAN}[3]{##3}\VANthebibliography}

\usepackage{graphicx}	
\usepackage{amsmath}	
\usepackage{xcolor}
\usepackage{multirow}
\usepackage{soul}

\definecolor{mlp}{rgb}{0.294, 0.612, 0.827}

\title[Isolating granulation]{Towards understanding stellar variability at the sub m/s level: Isolating granulation signals in synthetic spectral lines}

\author[G Frame et al.]{\parbox{\textwidth}{\Large
Ginger Frame,$^{1, 2}$\thanks{E-mail: ginger.frame@warwick.ac.uk}
Heather M. Cegla,$^{1, 2}$
Veronika Witzke,$^{3}$
Cis Lagae,$^{1, 2}$
Michael L. Palumbo III,$^{4}$
Sergiy Shelyag,$^{5}$
Christopher Watson,$^{6}$
Alexander Shapiro$^{3,7}$ 
}
\vspace{0.2cm}
\\
$^{1}$Department of Physics, University of Warwick, Gibbet Hill Road, Coventry CV4 7AL, UK\\
$^{2}$Centre for Exoplanets and Habitability, University of Warwick, Gibbet Hill Road, Coventry CV4 7AL, UK\\
$^{3}$ University of Graz, Institute of Physics, Universit\"atsplatz 5, 8010 Graz, Austria\\
$^{4}$Center for Computational Astrophysics, Flatiron Institute, 162 Fifth Avenue, New York, NY, USA\\
$^{5}$College of Science and Engineering, Flinders University, Tonsley Innovation District, 5042, South Australia, Australia.\\
$^{6}$Astrophysics Research Centre, School of Mathematics and Physics, Queen’s University Belfast, Belfast, BT7 1NN, UK \\
$^{7}$ Max-Planck-Institut f\"ur Sonnensystemforschung, Justus-von-Liebig-Weg 3, 37077 G\"ottingen, Germany\\
}

\date{Accepted 2025 April 15. Received 2025 April 09; in original form 2025 March 18}

\pubyear{\the\year{}}

\begin{document}
\label{firstpage}
\pagerange{\pageref{firstpage}--\pageref{lastpage}}
\maketitle

\begin{abstract}
Granulation in the photospheres of FGK-type stars induces variability in absorption lines, complicating exoplanet detection via radial velocities and characterisation via transmission spectroscopy. We aim to quantify the impact of granulation on the radial velocity and bisector asymmetry of stellar absorption lines of varying strengths and at different limb angles. We use 3D radiation-hydrodynamic simulations from \texttt{MURaM} paired with \texttt{MPS-ATLAS} radiative transfer calculations to synthesise time series’ for four \ion{Fe}{I} lines at different limb angles for a solar-type star. Our line profiles are synthesised at an extremely high resolution (R = 2,000,000), exceeding what is possible observationally and allowing us to capture intricate line shape variations. We introduce a new method of classifying the stellar surface into three components and use this to parameterise the line profiles. Our parameterisation method allows us to disentangle the contributions from p-modes and granulation, providing the unique opportunity to study the effects of granulation without contamination from p-mode effects. We validate our method by comparing radial velocity power spectra of our granulation time series to observations from the LARS spectrograph. We find that we are able to replicate the granulation component extracted from observations of the \ion{Fe}{I}  617~nm line at the solar disk centre. We use our granulation-isolated results to show variations in convective blueshift and bisector asymmetry at different limb angles, finding good agreement with empirical results. We show that weaker lines have higher velocity contrast between granules and lanes, resulting in higher granulation-induced velocity fluctuations. Our parameterisation provides a computationally efficient strategy to construct new line profiles, laying the groundwork for future improvements in mitigating stellar noise in exoplanet studies.

\end{abstract}




\section{Introduction}

Granulation refers to the observed stellar surface appearance in low-mass stars, characterised by bright, hot, up-flowing granules and cooler, down-flowing intergranular lanes. Understanding this process and its impact on the formation of stellar absorption lines is crucial for both studying stellar atmospheres and detecting exoplanets. Granulation influences absorption lines, introducing line-shape variations that can obscure the radial velocity (RV) signals caused by orbiting exoplanets. The granulation-induced variations result in RV shifts that can be of the order of $1~\mathrm{m~s^{-1}}$ \citep{book}, far above the current precision obtained by high-resolution spectrographs such as ESPRESSO \citep{espresso},  EXPRES \citep{expres} and NEID \citep{NEID}, and drowning the $9~\mathrm{cm~s^{-1}}$ signatures of Earth-like planets. 


Various techniques have been proposed to mitigate the impact of granulation in radial velocity observations. \cite{Dumusque} and \cite{Meunier2015} suggest observational strategies to average out stellar variability. Both studies conclude that granulation-induced RV rms can be reduced by averaging multiple observations spaced apart by hours. However there is some disagreement in the effectiveness of this method, with \cite{Meunier2015} finding that the strategy fails to reduce granulation rms RV to below the noise level of current instruments. See \cite{ceglareview} for a more in-depth description of current methods for granulation mitigation in exoplanet detection. 

In exoplanet characterisation via transmission spectroscopy, granulation induced line shape variations can imprint subtle asymmetries and depth changes in exoplanetary absorption features, leading to biases in retrieved atmospheric properties. \cite{Chiavassa2019} show the importance of using 3D models to properly account for the center to limb variations (CLV) of the stellar spectrum. See \cite{NASA} for a review of the effects of stellar contamination on transmission spectroscopy.

Numerous works have shown that the impact of stellar activity varies line-by-line as different atmospheric layers are probed \citep{moulla,  Cretignier2020, Dravins2023, Lafarga_2023, Dumusque_2018}. Valuable information can be gained by studying the line shape characteristics and granulation-induced velocities of individual stellar lines. By quantifying the introduced asymmetries and relating them to convective blueshifts we may be able to correct for the effects of granulation. Work such as this requires an extremely high spectral resolution in order to properly assess bisector asymmetries. 

\citet{grass1, grass2} detail a method to empirically synthesize high-resolution spectra ($R \sim 700,000$) based on solar disk-resolved observations. They quantify RV variability for 22 solar lines within wavelength range 525-630nm and argue that correlations with bisector asymmetry measures could be used to remove 25-35\% of granulation-induced noise. A limitation in this study, as pointed out in \cite{grass1}, is an inability to expand this method to non-solar type stars. If we demonstrate that we are able to conduct similar analysis through simulated data, we open up the opportunity in future work for the inclusion of other stellar types. To achieve this, we must simulate time-series for several different stellar lines across a range of line-formation parameters and wavelengths, and benchmark against observational studies. 

We aim to conduct a study with lines synthesized through 3D Hydrodynamic (HD) and eventually Magnetohydrodynamic (MHD) simulations. This builds upon the work of \cite{cegla2013, cegla2018, cegla2019}, hereafter the Cegla Series (CS), who synthesized disk-integrated line profiles of the \ion{Fe}{I}  6302 \AA\ line for a solar type star with a 200~G magnetic field. CS used box-in-a-star MHD simulations of the stellar surface from \texttt{MURaM} \citep{MURaM}, paired with \texttt{NICOLE} \citep{nicole1, nicole2} for line synthesis. Since CS was published, improvements have been made to the \texttt{MURaM} code: 
\citet{rempel2014} modified the diffusion scheme to reduce the diffusive terms as much as possible, while ensuring numerical stability. To achieve a better agreement with solar observations, the new \texttt{MURaM} setup uses the equation of state look-up-table generated by the FreeEOS code \citep{Irwin_freeeos_2012}, together with an updated and consistent opacity table using the element composition by \citet{Asplund_2009}.  Furthermore, the radiative treatment was upgraded to consider 12 multi-group bins with similar thresholds as in \citet{Magic_2013A&A}. For details on how these changes affect the accuracy of the simulations, see Appendix A.3 and B in \citealt{witzke2024}.


The goal of this work (which will be over a series of papers) is to continue the CS studies by using a wider range of atomic absorption lines and various magnetic field strengths whilst utilising the improved \texttt{MURaM} model atmospheres. The end-goal of this effort is to produce accurate, high resolution, disk-integrated simulated observations in spectral lines for differing stellar types with different magnetic field strengths. We aim to quantify the effect of magnetic fields on granulation-induced line-shape changes in future work. In this paper, we first analyse granulation effects on disk-resolved spectral lines in the hydrodynamic (HD) case, laying the groundwork for future investigations of magnetic fields. We use this paper to introduce the required methods that will remain applicable throughout future work. 


To isolate the granulation effects from the pressure-modes present in \texttt{MURaM} MHD cubes and also to enable the reconstruction of lines with a lower computation cost, CS uses a method of parameterisation (see \cite{cegla2013} for details). In this paper, where we focus on the HD case, we find that parts of this method are insufficient to properly remove the effects of p-modes whilst maintaining the granulation signature in our data. We attribute this to a combination of updates to MURaM and the much stronger p-modes present in the HD case compared to a 200G magnetic field.

We introduce a more rigorous parameterisation approach to model the impact of granulation on stellar line profiles. This method leverages information from 3D HD simulations to generate line profiles without requiring the computationally expensive step of rerunning the full simulations. By doing so, it significantly reduces computation time while retaining the critical details provided by the simulations. 

Traditional approaches to modelling spectral lines with 1D hydrostatic stellar atmosphere models rely on approximate methods to account for the missing large- and small-scale velocity fields. These methods rely on free parameters such as micro- and macro-turbulence, which act as simplified 1D approximations of complex 3D turbulent flow structures present in stellar atmospheres, despite their known inaccuracies \citep{Yoichi, Han}. In contrast, the parameterisation approach incorporates the complex convection processes directly encoded in the parameters derived from the HD simulations, providing a simpler and more accurate way to account for the full range of stellar atmosphere information whilst reducing computation time. We also preserve granulation time-variability with our approach, allowing us to retain information on the evolution of granulation. 

A description of the data used in this work is given in Section \ref{sec:data}. An outline of our parameterisation method and a demonstration of its applicability at different positions on the stellar disk is provided in Section \ref{sec:para}. A description of how the time series is reconstructed and comparison to observed power spectra are presented in Section \ref{sec:recon}. Section \ref{sec:limb} provides an analysis of center-to-limb effects for the different lines. Finally, Section \ref{sec:sum} contains a summary and conclusions from the study.

\section{Data} \label{sec:data}

\begin{figure}
    \centering
    \includegraphics[width=\columnwidth]{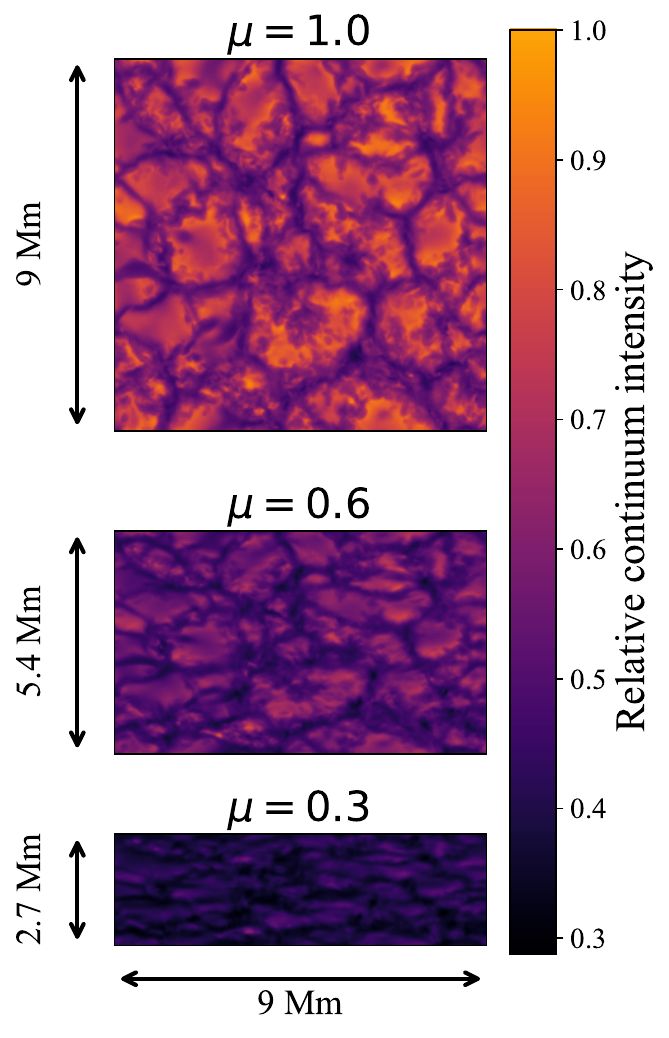}
    \caption{The continuum intensity at 617~nm plotted for each pixel at three limb angles. The units of intensity are arbitrary and scaled relative to the maximum intensity at disk center. Projected spatial dimensions of the cube at each limb angle are indicated with double arrows.}
    \label{fig:cont}
\end{figure}

\begin{table*}
    \centering
    \begin{tabular}{l c c c c c c c c}
    \hline
    Element + ionisation & Air wavelength (nm) & log(\textit{gf}) & Land\'e factor& $\gamma_{\text{rad}}$ & $\gamma_{\text{Stark}}$ & $\gamma_{\text{VdW}}$ & Excitation potential (eV) \\
    \hline
    \ion{Fe}{I}  & 525.021 & -4.938$^{\dagger}$  & 3.00    & 3.32  & -6.28 & -7.82$^*$ & 0.121 \\
    \ion{Fe}{I}  & 615.162 & -3.299$^{\dagger}$ & 1.84 & 8.29 & -6.16 & -7.70$^*$ & 2.176 \\
    \ion{Fe}{I}  & 617.333 & -2.880$^{\dagger}$  &  2.50  & 8.31  & -6.16 & -7.69$^*$ & 2.223 \\
    \ion{Fe}{I} & 627.128 & -2.701$^{\ddagger}$ &  1.49 & 8.23 & -5.41 & -7.28$^*$ & 3.332 \\
    \hline
    \end{tabular}
    \caption{Atomic parameters for the lines selected for this work. $\gamma_{\text{rad}}$, $\gamma_{\text{Stark}}$ and $\gamma_{\text{VdW}}$ refer to the radiative, Stark and Van der Waals damping parameters. All data is sourced from \protect\cite{K14}, unless marked otherwise. \\[0.5ex]
    \textbf{Key:} $^\dagger$ Source: \protect\cite{FMW}; $^\ddagger$ Source: \protect\cite{BK}; $^*$ Source: \protect\cite{BPM}.}
    \label{tab:atomic_data}
\end{table*}

\begin{figure*}
    \centering
    \includegraphics[width = \textwidth, trim=0cm 0cm 0cm 0cm, clip]{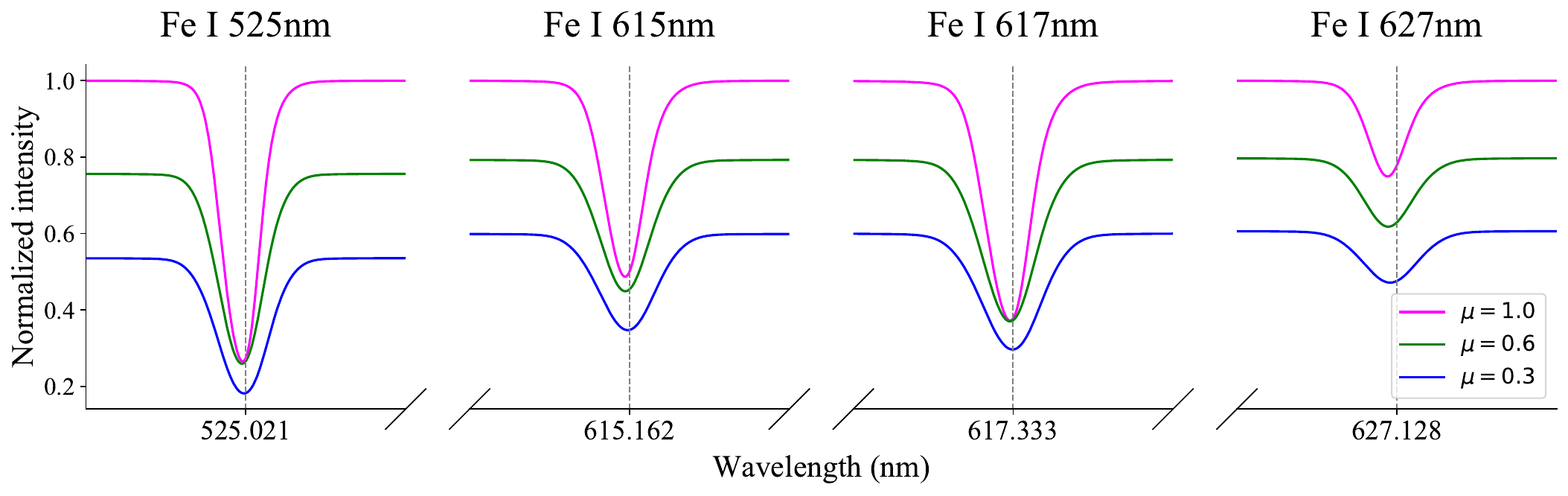}
    \caption{Spatially and temporally averaged line profiles calculated using the \texttt{MURaM} cubes with \texttt{MPS-ATLAS} at three limb angles $\mu=1.0,~0.6$ and $0.3$. The intensity of each profile has been normalised relative to the continuum intensity of the relevant disk-centre profile. The gray dotted lines are the rest wavelengths in air of the lines. Each section separated by the double diagonal lines covers the wavelength interval of 0.04~nm centered at the rest wavelength.}
    \label{fig:lps}
\end{figure*}

\begin{figure*}
    \centering
    \includegraphics[width = \textwidth, trim=0cm 0cm 0cm 0cm, clip]{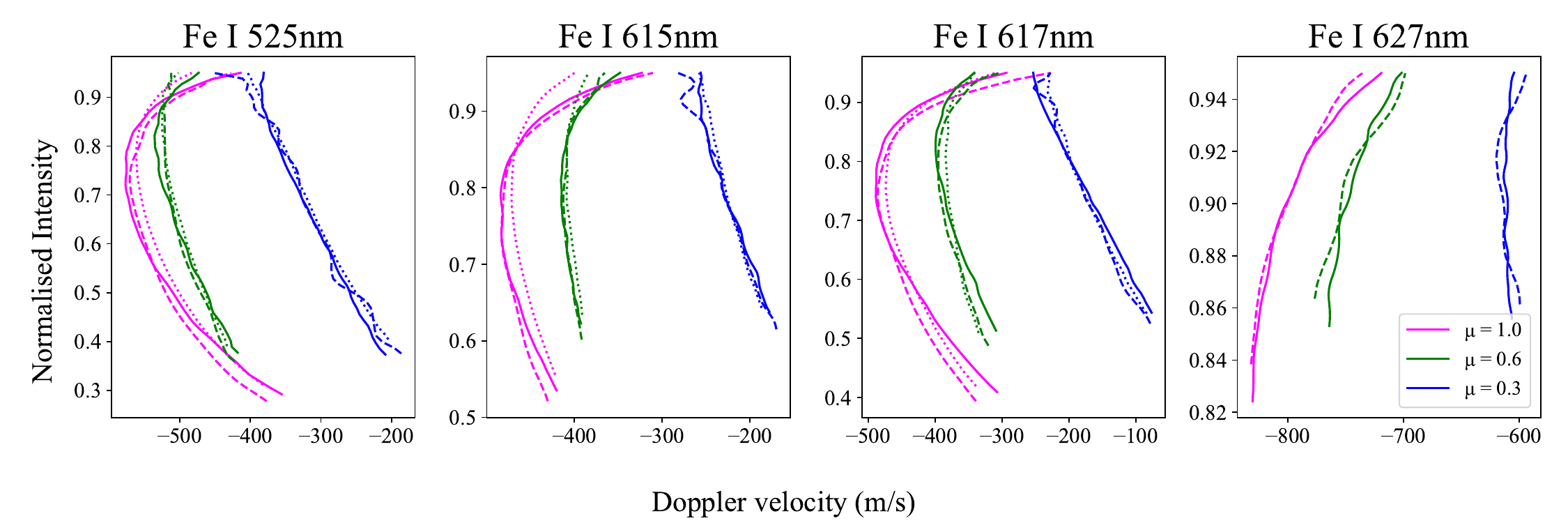}
    \caption{Bisectors of the profiles displayed in Figure \ref{fig:lps}. Intensity has been normalised relative to the relevant continuum intensity at each inclination. Note that each subplot has an independent y-axis scale to better visualize variations. Solid lines are results from this work, dotted lines are bisectors from LARS and dashed lines are bisectors from IAG. Both the LARS and IAG bisectors have been shifted in velocity space to match the mean bisector Doppler shift of the synthesised profiles.}
    \label{fig:bis}
\end{figure*}

We use a time series of 3D HD models generated using \texttt{MURaM} (\citealt{MURaM}) for a solar-like star ($T_\mathrm{eff}$ = 5787 K) . Each simulation box has a horizontal extent of of 9 $\times$ 9~Mm$^2$ containing $512\times512$ pixels, and a vertical extent of 5~Mm containing 500 pixels. In the vertical direction, the model covers approximately 1~Mm of atmosphere above the photosphere and 4~Mm of sub-photosphere and convection zone, and is set on a uniform mesh with a spacing of 10 km. Our time series consists of 129 snapshots with a cadence of 30 seconds, spanning a total time of just over 1 hour. 

As an artifact of the model initialisation and due to periodic side boundaries, the HD cubes show a horizontal drift through their periodic boundary conditions. To diminish the impact of this artificial velocity field on our results, we subtract the average horizontal root mean square (rms) velocity of the cube at each timestep. We observed no change in our parameterisation results after this, due to the relative pixel-to-pixel velocity shifts being untouched. 

For each of the $512 \times 512$ vertical columns in each snapshot, we synthesise line profiles using \texttt{MPS-ATLAS} \citep{mps-atlas}, a 1D radiative transfer code that assumes local thermodynamic equilibrium (LTE). Line profiles are synthesised in the so-called 1.5D approach, meaning that the velocity fields of the 1D columns are inherited from the 3D cube. As such, no microturbulence is required for these calculations. Our line profiles have a resolution of $R = 2,000,000$.  We choose to synthesise four lines at nine different limb angles ranging from $\mu = 0.2$ to $\mu = 1.0$ in steps of 0.1, and only one azimuthal angle. Note that $\mu$ in this context refers to the cosine of the heliocentric angle, with $\mu=1.0$ corresponding to the disk centre. Before each line synthesis calculation, the hydrodynamic quantities of the 3D cubes are tilted and interpolated onto rays parallel to the viewing angle. The resulting vertical resolution of the new tilted cubes is kept similar to the original cube. For a more detailed description of how the \texttt{MURaM} boxes are generated and how \texttt{MPS-ATLAS} line synthesis works, refer to Section 2 in \cite{witzke2024} and references therein.

Figure \ref{fig:cont} presents an example of a single snapshot in time observed at three different heliocentric angles. The continuum intensity values displayed in this plot have been normalised by the maximum intensity at disk centre. Continuum intensity decreases towards the limb due to limb darkening. A consequence of inclining the cubes is a change in which pixels remain visible to the observer. The optical surface is located at higher altitudes in granules and lower in lanes, creating a 3D corrugated surface structure. By viewing the cube at an inclined angle, some pixels become obscured behind others. See \cite{Dravins2008} for further discussion.

The four lines we synthesise are \ion{Fe}{I}  617~nm, \ion{Fe}{I}  525~nm, \ion{Fe}{I}  627~nm and \ion{Fe}{I}  615~nm. The focus on \rm{\ion{Fe}{I}} lines for this work is due to their sensitivity to photospheric conditions and largely well-established atomic properties. These four lines span a range of different line strengths, Land\'e factors and excitation potentials. Their atomic properties can be found in Table \ref{tab:atomic_data}. Note that the Land\'e factors will have no effect on the results of this work, since we are not including a magnetic field in our model atmosphere. However these values will become important in future comparative studies including magnetic fields.

Figure \ref{fig:lps} shows the spatially and temporally averaged \texttt{MPS-ATLAS} output for each spectral line at three limb angles. As before, the intensity values have been normalised relative to the continuum intensity at disk centre. The gray dotted lines indicate the rest wavelength of each line. Note that due to the averaging, these lines contain no p-mode information and the evident convective blueshift seen in Figure~ \ref{fig:lps} is a direct result of stellar surface granulation, where hot up-flowing granules occupy a larger surface area fraction compared to intergranular lanes \citep{Dravins1981}. A more in-depth discussion on line shape changes across different limb angles for the different lines is given in Section \ref{sec:limb}.

In Figure \ref{fig:bis}, we compare the bisectors of the averaged line profiles from Figure \ref{fig:lps} with disk-resolved observations from the Laser-based Absolute Reference Spectrograph (LARS) \citep{LARS} and from the Institut für Astrophysik, Göttingen (IAG) solar atlas \citep{IAG}. The intensity is normalized relative to the continuum intensity at each $\mu$. Doppler shifts are computed using the rest wavelengths listed in Table \ref{tab:atomic_data}. However, a direct comparison of the simulated and observed mean bisector Doppler shift is not meaningful due to potential large-scale flows in the observed data (e.g., supergranulation, meridional flows). Instead, we focus on comparing the bisector shapes and align the observed profiles by shifting them to match the mean bisector velocity of our results. In Figure \ref{fig:bis}, the dotted lines are the shifted LARS bisectors (if available) and the dashed lines are the shifted IAG bisectors. We find a sufficient match in all cases, any differences can be attributed to the exclusion of magnetic effects in our simulations and the assumption of LTE in our line synthesis. It is also important to note that LARS and the IAG FTS do not match the spectral resolution of our synthetic spectra, and both are affected by an instrument-specific line spread function (LSF), which remains uncharacterised in this analysis.

\section{The Parameterisation method}
\label{sec:para}

\begin{figure}
    \centering
    \includegraphics[width=\columnwidth]{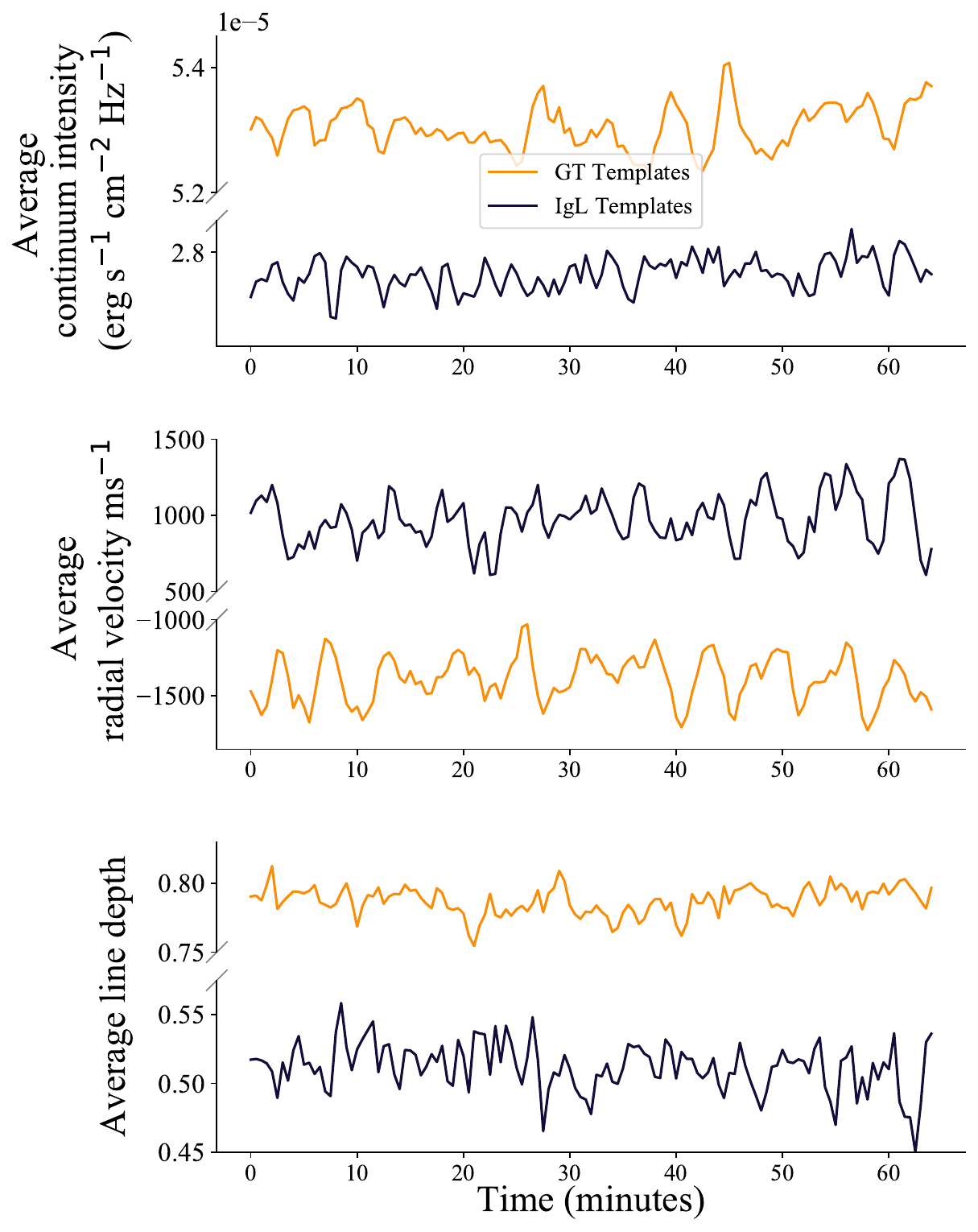}
    \caption{The evolution of averaged parameters for each initial template group with time. Template groups are defined as the brightest and darkest 2\% of pixels in each snapshot. The bright templates represent Granular Tops (GTs) and the dark templates represent Intergranular Lanes (IgL). The periodic instances seen here are an artifact of p-modes. Radial velocity is calculated by cross correlating the line profile from each pixel with the total temporal and spatially averaged line profile.}
    \label{fig:templates}
\end{figure}

\begin{figure*}
    \centering
    \includegraphics[width=\textwidth]{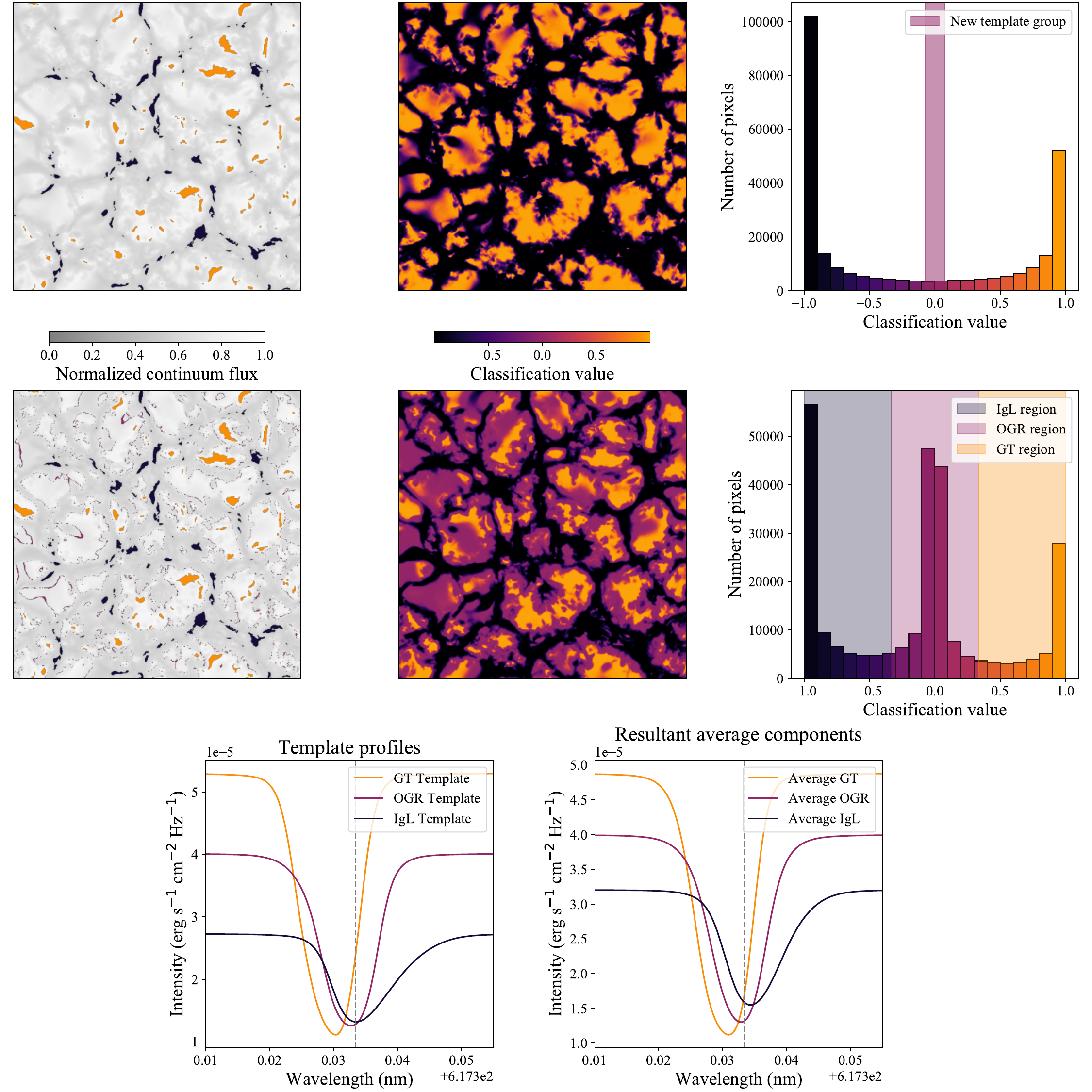}
    \caption{The parameterisation method demonstrated on a single snapshot, using the \ion{Fe}{I}  617nm line. \textit{Top row:} The top left corner plot shows the locations of the Granular Top (GT) template pixels in orange and the Intergranular Lane (IgL) template pixels in black, overlaid on the normalised continuum flux. To the right of this are the resulting classification values for every pixel from the logistic regression model trained on the template pixels. Highlighted in purple in the top right plot is the region from which Outer Granular Region (OGR) template pixels are selected. \textit{Middle row:} The leftmost plot again shows the locations of the templates, this time including the OGR template group in purple. To the right of this are the final classification values determined by the retrained logistic model including all three components. \textit{Bottom row:} The bottom two plots show the spatially averaged template line profiles on the left and the resultant spatially averaged components on the right. The grey dotted lines indicate the rest wavelength of the line.}
    \label{fig:class}
\end{figure*}

\begin{figure*}
    \centering
    \includegraphics[width = \textwidth]{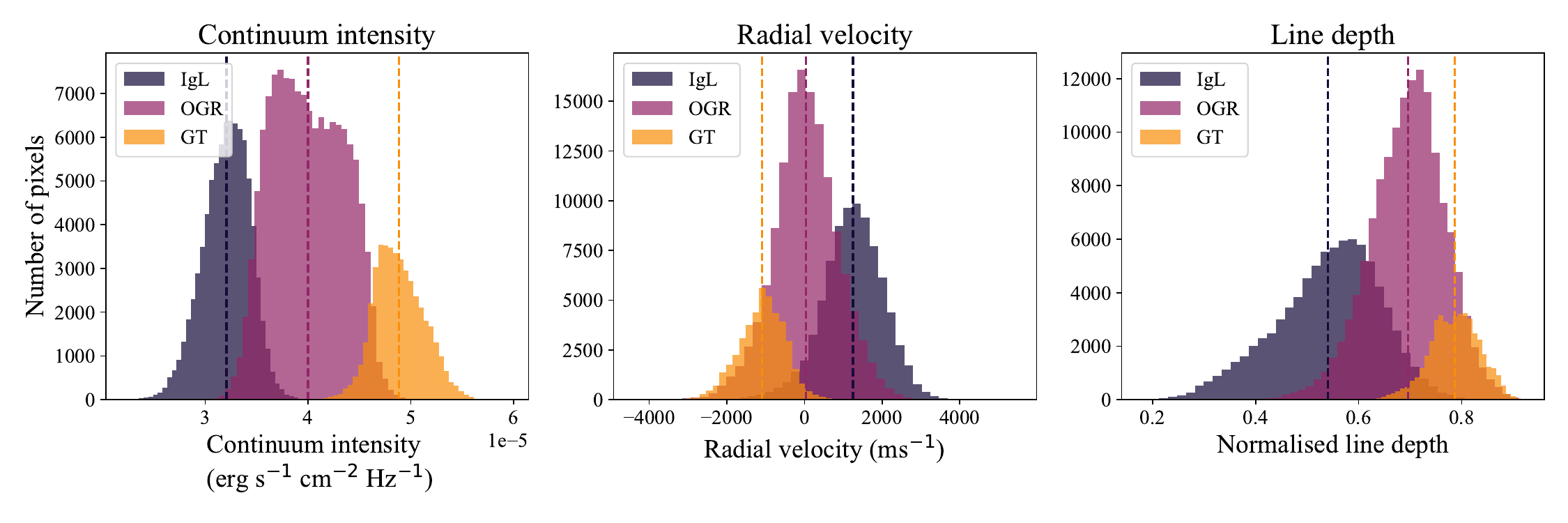}
        \caption{Distributions of selected parameters for each component for one snapshot at disk centre (\(\mu = 1.0\)). The dotted lines indicate the mean value of each component. Note that radial velocity here is calculated relative to the time averaged profile across all snapshots (which itself exhibits a convective blueshift).}
    \label{fig:hist}
\end{figure*}

Generating line profile time series from HD cubes, as described in Section \ref{sec:data}, is computationally demanding. Constructing disk-integrated profiles requires tiling a stellar grid (see \citealt{cegla2019}), but covering a solar-sized star with unique 9~Mm$^{2}$ HD cubes is infeasible due to computational costs. While repeating cubes across multiple tiles is possible, it may introduce biases in the resulting disk-integrated profile. Instead, we propose a new parameterisation method that leverages a single time series of HD cubes to construct unique, accurate line profiles, eliminating the need for separate HD simulations for each tile.

As well as computational efficiency, parameterisation allows us to disentangle the line changes driven by p-modes and granulation. This offers us a valuable opportunity to study the impact of granulation completely uncontaminated by other variability processes. 

The idea of separating line profiles into distinct surface components was pioneered by \cite{Dravins1990}. In our approach, the HD cube is seperated into components such as granules and intergranular lanes. We assume that each component has a distinct line-profile contribution, and solve for this contribution by averaging profiles from `granule' and `lane' pixels both spatially and temporally. We then assume that granulation-induced line shape changes can all be attributed to changes in component filling factor, which refers to the changing proportion of pixels belonging to each component in each snapshot. For an example of this method in practice, see \cite{cegla2013} and the figures therein.


The difficulty of this process lies in defining components and classifying pixels in each snapshot. In \cite{cegla2013}, cutoff values are defined in both continuum intensity and magnetic field strength. These cutoffs are applied uniformly to every snapshot to separate pixels into four components. However, in our non-magnetic dataset, we find that the average continuum intensity of each snapshot is heavily influenced by p-mode phase. Using a universally applied continuum cutoff value results in filling factors contaminated by the periodic effects of p-modes. We therefore present an updated technique for parameterisation that is more careful to ensure the elimination of p-mode effects whilst maintaining granulation signatures. This technique is applied independently to each of the four lines from Table~\ref{tab:atomic_data}. However, for the purpose of demonstrating this method, we will use the \ion{Fe}{I} 617~nm line for the entirety of this section. 


\subsection{Pixel classification}

It is vital to ensure that the parameters we use to assign pixels to specific components are not influenced by p-modes, so that the averaged components we end up with are free from contamination. This is challenging because both granulation and p-modes have overlapping effects on all available parameters. However, a key distinction is that granulation primarily influences the spatial characteristics of the cube, while p-modes mainly affect the temporal aspects: at any given moment, the pressure mode amplitude in the cube is mostly constant across pixels, while granulation causes spatial variability between the pixels. By examining a single snapshot, we can observe this spatial variation in granulation effects without the interference of time-dependent p-mode fluctuations. Therefore, when classifying pixels, we compare them only to other pixels within the same snapshot to determine their component. 

We separate the pixels in each snapshot into three distinct groups, partially motivated by \cite{Dravins1990}, who also chose to treat granules separately from their centers.

\begin{itemize}
    \item Granular Tops (GTs): the brightest, hottest points on a granule in which velocity up-flows are strongest. 
    \item Outer Granular Regions (OGRs): The area of the granule surrounding the GT.  These are a mixture of up-flows and down-flows as these regions contain plasma both rising to the GT and falling into the lanes. 
    \item Intergranular Lanes (IgLs): The cooler regions that the plasma sinks into, consisting of mostly down-flows. 
\end{itemize}

Each pixel in every snapshot is assigned to one of these groups. We base this decision on three key spectral parameters: the continuum intensity, the measured radial velocity, and the line depth. These chosen parameters capture essential aspects that track variability in the line and surrounding spectral region whilst minimizing potential biases from parameter correlations. All radial velocities in this work are calculated by cross-correlating the line profiles with the total time averaged profile as the template. We fit a parabola to the peak of the cross-correlation function (CCF) to determine the RV. 

However, each of these parameters is contaminated snapshot-to-snapshot by p-mode oscillations. We therefore need to produce template parameters that evolve with the snapshots, thus capturing what each group should look like at a particular phase in the p-mode cycle. 

We begin with a simple assumption that the very brightest points in a snapshot are GTs and the very darkest points are IgLs. We therefore use parameters from the brightest 2\% of pixels as an initial GT template and the darkest 2\% as an initial IgL template. Figure \ref{fig:templates} shows how the average of these template parameters evolves with time. Note the instances of periodic behavior caused by p-mode oscillations. See the top leftmost plot in Figure \ref{fig:class} for locations of the template pixels.

A logistic regression model is trained on the template parameters for the two groups. This was done using scikit-learn's LogisticRegression class \citep{scikit-learn}. This model takes into account the full spread of parameters given and inherently assigns feature importance values for each parameter based on how strongly it correlates with the grouping. For each of the $512\times512$ pixels, the probability of belonging to the GT group (P(GT)) or the IgL group (P(IgL)) is estimated by the model based on the pixel parameters compared to the template values. The classification value is then determined as follows, resulting in a range from -1 (definitely a GT) to +1 (definitely an IgL). 
\hfill \break

\text{Initial classification} = 
$\begin{cases} 
    \frac{\text{P(GT)} - 0.5}{0.5} & \text{if } \text{P(GT)} > \text{P(IgL)} \\[2ex]
    -\frac{\text{P(IgL)} - 0.5}{0.5} & \text{otherwise}
\end{cases}$
\hfill \break
\hfill \break
\hfill \break

The initial classification results can be found in the first row of Figure \ref{fig:class}. Note that at this stage we are only considering GTs and IgLs and are yet to introduce template parameters for the OGR group. We do this now by considering the 2\% of pixels that have a classification value closest to zero (see the top rightmost plot in Figure \ref{fig:class}). These pixels have been determined by the trained model to have an equal probability of being GTs or IgLs. By using these pixels as our templates for the OGR region group, we ensure that our three groups are as distinct as possible. We retrain the logistic regression model with the inclusion of this third template group and calculate again P(GT) and P(IgL). This time we also calculate the probability that the pixel is an OGR (P(OGR)). Note that the sum of these three values for each pixel will sum to 1. We can therefore define our final classification value as 
\hfill \break

\text{Final classification} = P(GT) - P(IgL)
\hfill \break
\hfill \break
This again results in a range from -1 (definitely an IgL) to 0 (definitely a OGR) to 1 (definitely a GT). Pixels are then assigned a group based on their classification value. GTs have values between 1/3 and 1, OGRs between -1/3 and 1/3, and IgLs between -1 and -1/3. The final classification results are shown in the second row of Figure \ref{fig:class}.  The bottom row  of Figure \ref{fig:class} shows the average line profiles of the template pixels along with the resultant averaged group components. These spatially averaged group components are calculated for each snapshot individually and then are averaged again over time to remove the effects of p-modes. 

The decision to populate each of our template groups with 2\% of the available pixels is somewhat arbitrary. We tested the process also using 3\% and 5\%, and whilst there is a difference in initial classification result, as long as the percentage used is matched in the third template group, there is minimal difference in the determined final classification values. 

Figure \ref{fig:hist} shows the distributions of the chosen key line-shape parameters for each group in one snapshot. It is evident here that the classification method successfully separates all three parameters into groups that are as distinct as possible, prioritising continuum intensity since this parameter has the highest feature importance. 

The entire classification process is run separately for each of the four lines, at each $\mu$. Note that a pixel assigned to one group for one line may not necessarily be of the same group for a different line due to changing formation heights and differing contrast levels between granules and lanes.

\subsection{Pixel classification at different inclinations}
\label{sec:diff_inc}

\begin{figure}
    \centering
    \includegraphics[width=\columnwidth]{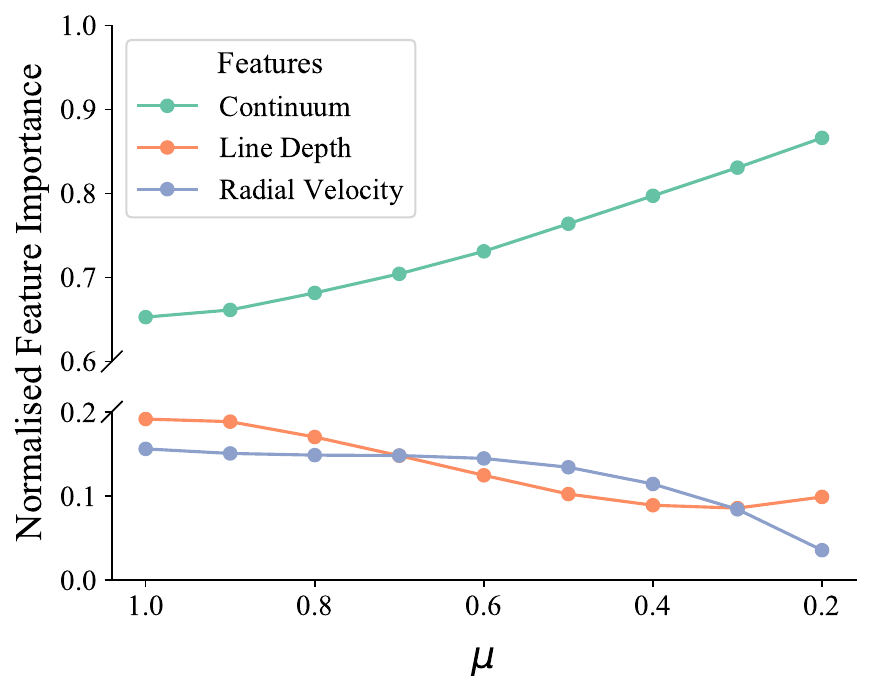}
    \caption{Average normalised absolute feature importance values at different limb angles. Each value has been divided by the sum of feature importance values to demonstrate the relative contributions of each feature and averaged over time. A high feature importance value indicates a strong weighting for the feature in the classification model.}
    \label{fig:importance}
\end{figure}

The classification method remains robust across different inclinations, as the templates are data driven and inherently adapt to the specific characteristics of each cube. By defining GT and IgL templates directly from the data at each inclination, the method ensures that classifications are accurately aligned with the unique distribution of line shape parameters. 

Figure \ref{fig:importance} illustrates the variation in feature importance (FI) values assigned by the logistic regression model at different inclinations. The FI value determines the relative weight of each feature in the classification of individual pixels. A higher FI indicates a stronger correlation between the feature and the classification outcome. Each point in the plot represents the absolute value of the FI for a feature, normalized by the sum of all FI values, and averaged over time. This reveals the relative contribution of each feature to the classification decisions. It is clear that continuum intensity consistently plays the dominant role in pixel classification, with this influence becoming more pronounced towards the limb. As higher atmospheric layers are probed, and horizontal velocity flows are incorporated, the distinction between the velocities and line depths of different components diminishes. Therefore the values become less important for classification purposes.

\begin{figure}
    \centering
    \includegraphics[width = \columnwidth]{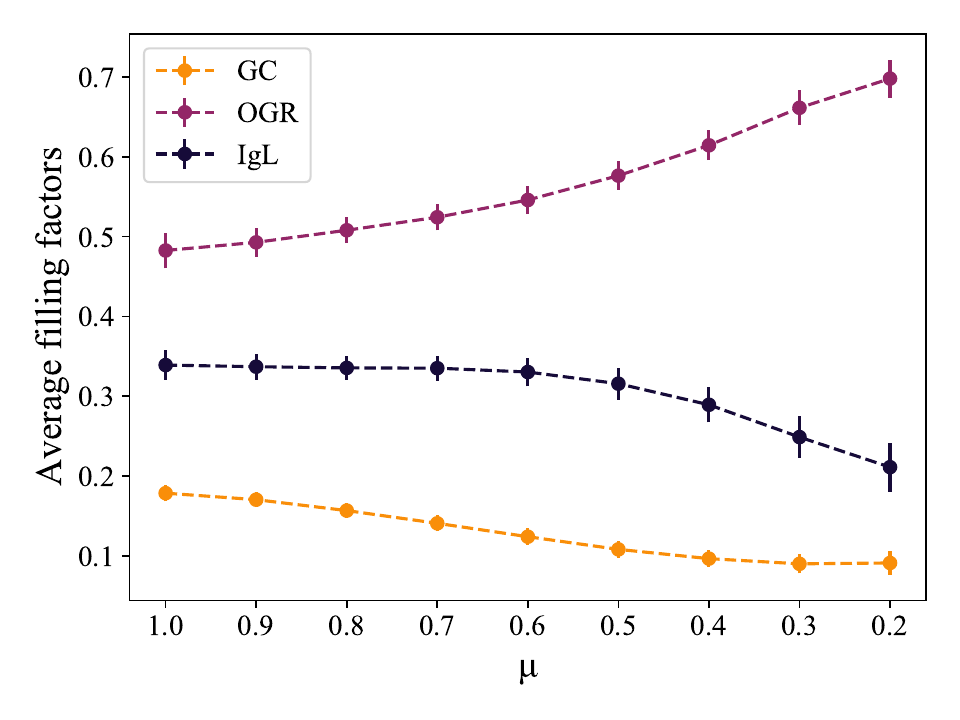}
    \caption{Time averaged filling factors of each component against limb angle for \ion{Fe}{I}  617~nm. The error-bars represent the standard deviation of each result.}
    \label{fig:ffs_limbs}
\end{figure}

Another consequence of viewing the cube from an inclined angle is a change in component filling factors. Figure \ref{fig:ffs_limbs} shows the average filling factor for each component at each limb angle. Note that as expected, we see an increase in OGR filling factor as we get closer to the limb due to extremities (IgLs and GTs) being concealed by the edges and peaks of the corrugated optical surface. 

\begin{figure*}
    \centering
    \includegraphics[width = \textwidth, trim=0cm 0cm 0cm 0cm, clip]{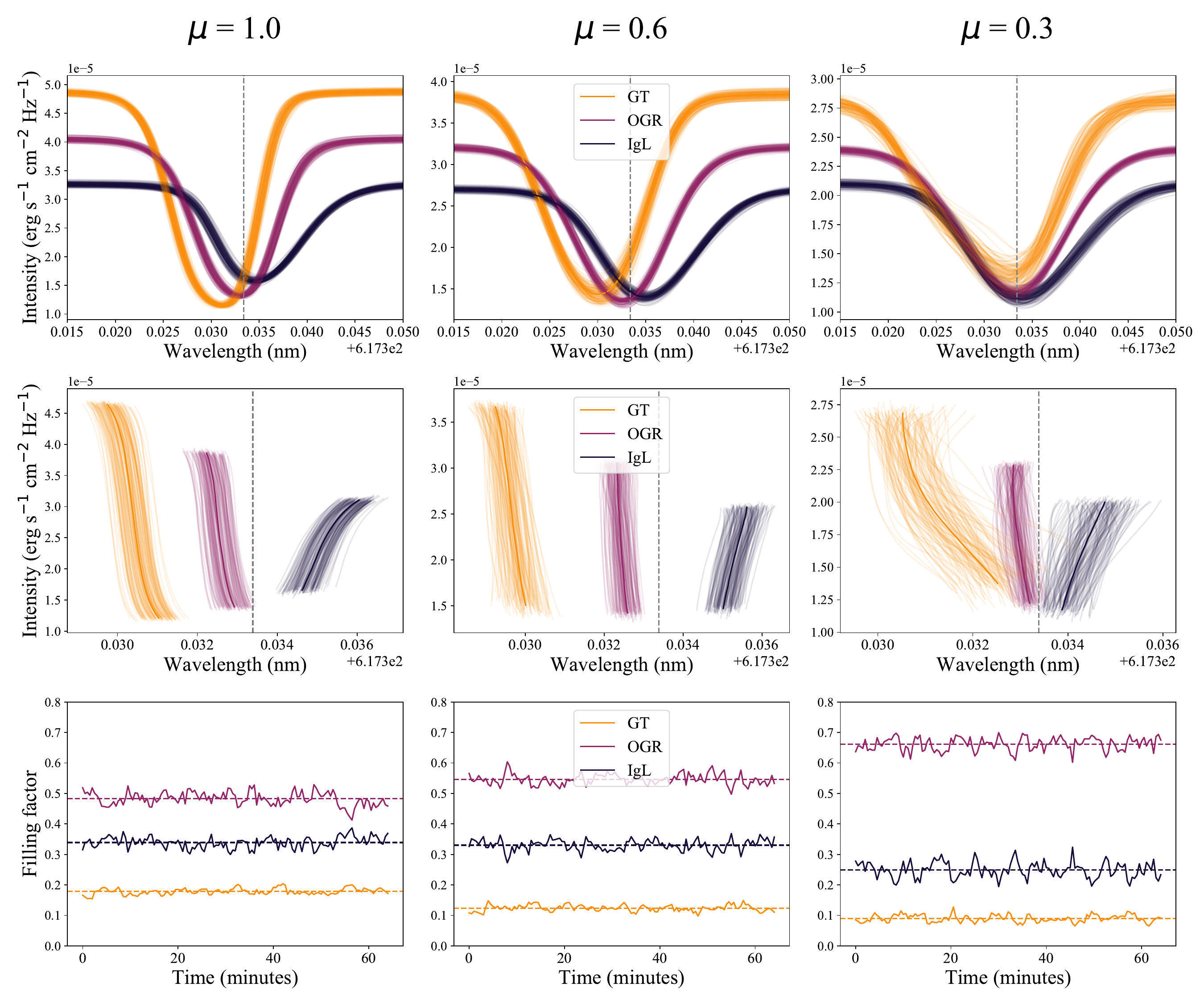}
    \caption{The parameterisation results for \ion{Fe}{I}  617~nm. \textit{Top row:} The spatially averaged components for every snapshot plotted on top of one another. \textit{Middle row:} The bisectors of the profiles above, with the bold lines indicating the time averaged result. Vertical gray dotted lines indicate the rest wavelength of the line. \textit{Bottom row:} The evolution of filling factors for each component over time. The horizontal dotted lines indicate mean values.}
    \label{fig:inc_res}
\end{figure*}

The results for the spatially averaged GT, OGR and IgL components are plotted in Figure \ref{fig:inc_res} at three limb angles, along with a time series of the filling factors. Scatter in velocity and intensity seen in the top two panels of this figure are a result of the p-modes. The reduction in contrast between component features can be clearly seen here in the profiles and bisectors at $\mu$ = 0.3. More discussion on changes to granule/lane contrast towards the limb can be found in Section \ref{sec:limb}.

\section{Reconstructing the time series}
\label{sec:recon}

\begin{figure*}
    \centering
    \includegraphics[width = \textwidth]{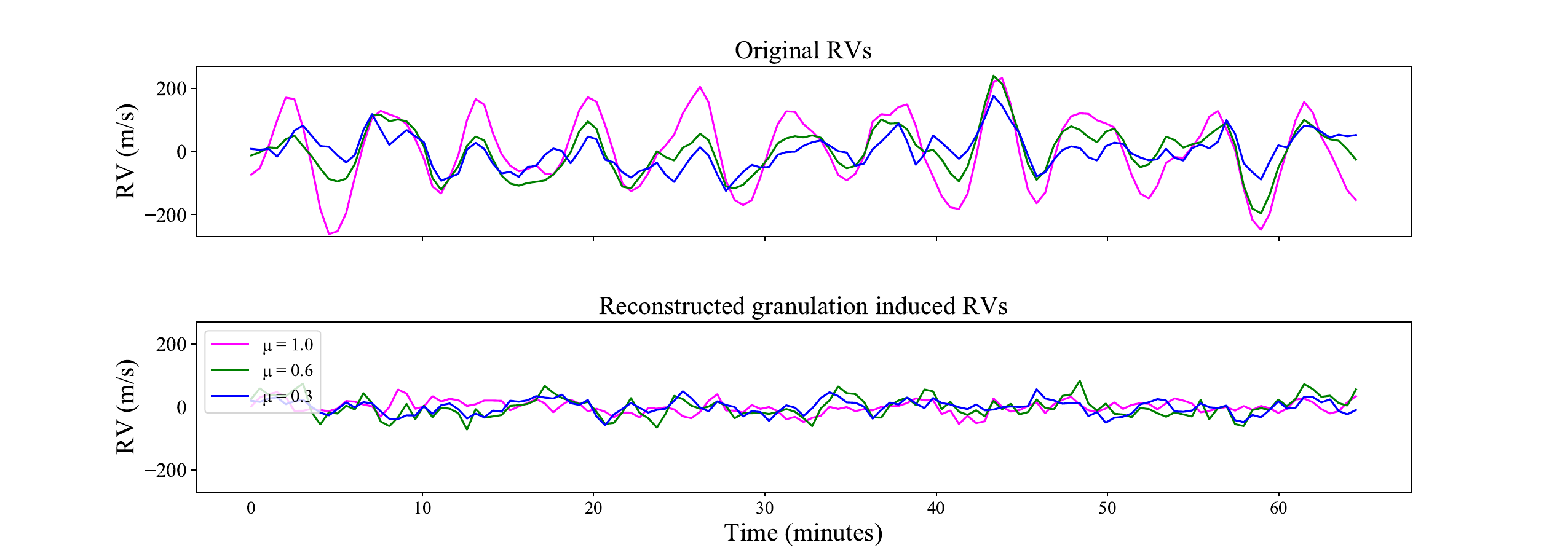}
    \caption{Line of sight radial velocities calculated by cross-correlating each profile with the time-averaged profile at the relevant limb angle. The top plot shows velocities from the original dataset that includes contributions from both granulation and p-modes. The bottom plot shows velocities from the reconstructed dataset, in which the p-modes have been removed.}
    \label{fig:rvs}
\end{figure*}

We reconstruct our line profiles using the time averaged component line profiles coupled to the component filling factors obtained from each HD snapshot. As previously discussed in Section \ref{sec:para}, the time averaging of components will eliminate the effects of p-modes, whilst the granulation signature is retained within the filling factors. Therefore by performing this reconstruction, we recreate the original time series but without p-mode effects, thus isolating the impact of granulation. Figure \ref{fig:rvs} shows the original RVs that include effects from both p-modes and granulation, and the RVs from the reconstructed time series, that include only granulation induced effects. As a reminder, RVs in this work are calculated using cross-correlation with a time averaged profile as the template. Figure \ref{fig:rvs} shows that the periodic oscillations caused by p-modes have been removed in the reconstructed time series.

\subsection{Validation via power spectra comparisons with observations}
\label{sec:power}

\begin{figure}
    \centering
    \includegraphics[width = \columnwidth]{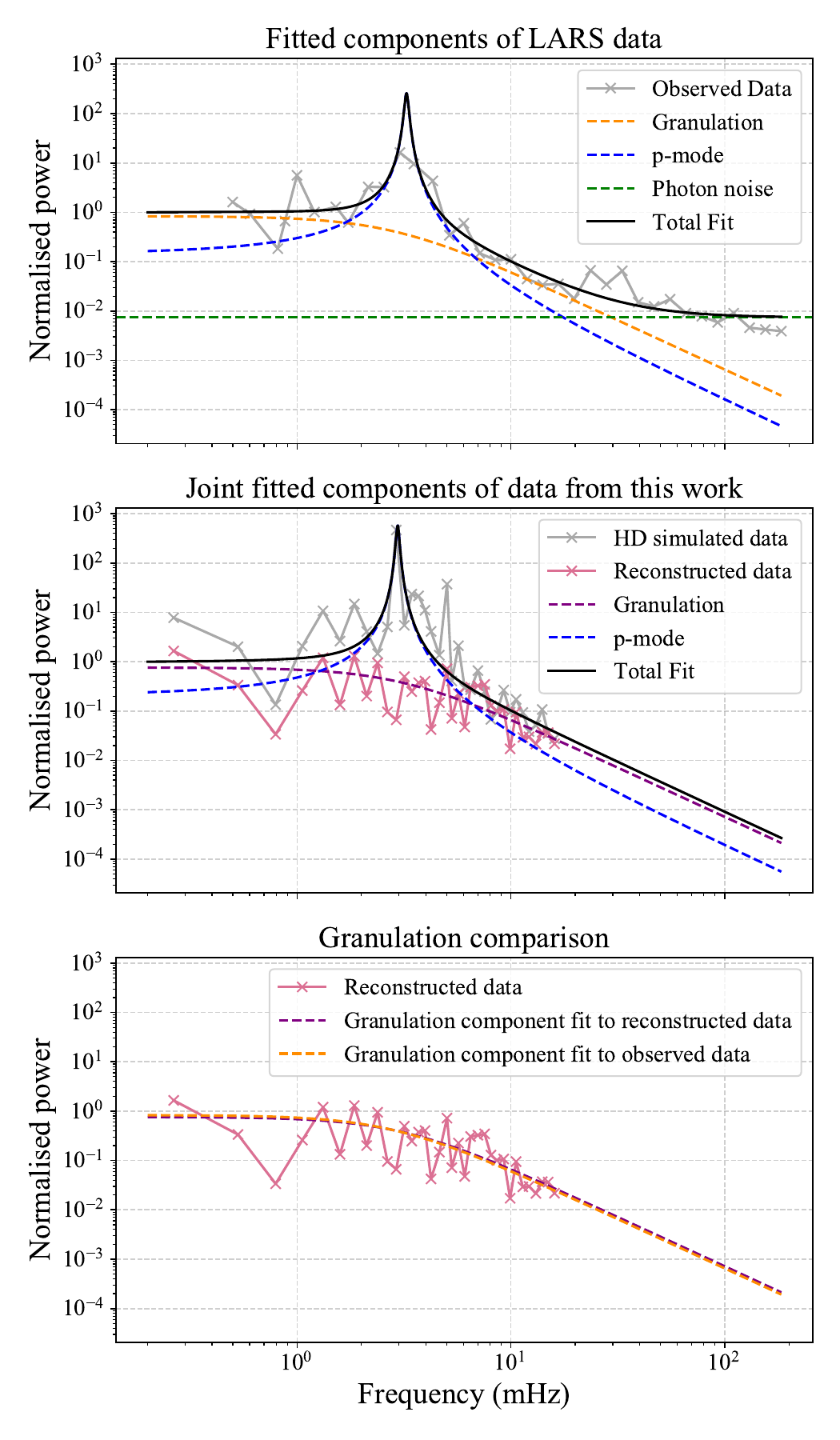}
    \caption{Velocity power spectral density and fits to simulated and observed datasets. The bold, black lines show the total fits to the data, while the dotted lines show the components within this fit. The top plot displays observational data from LARS. The middle plot shows both the original HD and reconstructed versions of our simulated dataset. In this case, a joint fit was performed, with the assumption that the reconstructed data represented the granulation component of the original data. The bottom plot shows the reconstructed data from our simulations with both granulation components from the plots above.}
    \label{fig:power}
\end{figure}

Recent work by \cite{al2023stellar} separated stellar signal components by fitting analytical functions to radial velocity power spectral densities (VPSDs; \citealt{lefebvre2008variations}). Other examples of this method can be found in \citealt{cegla2018}, \citealt{Kallinger2014} and \citealt{Michel2009}. Here, we use similar techniques to validate our reconstruction method.

Our HD simulations include p-mode oscillations and granulation. If the reconstruction using the parameterised line profiles is successful, the reconstructed time series should exclude p-mode contributions while preserving the granulation signature. We characterize the granulation contribution to the VPSD and validate our reconstruction method against observations of the solar VPSD, described below. 

The observation and data-reduction procedures for the LARS data used in this section are described in greater detail in \citet{Lohner-Bottcher2017} and \citet{lohner2019}. Here, we use eight time series observed at solar disk center that are centered on the \ion{Fe}{I}  617~nm line. In the analysis presented in \citealt{grass2}, these data were binned to a 15-second cadence; here we use the unbinned data that were observed at a 1.5-second cadence (0.5 seconds exposure + 1.0 second overhead). The baselines of the time series vary somewhat, but the longest covers $\sim$29 minutes and the shortest $17$ minutes. We calculate separate VPSDs for each of the eight time series, from which we calculate a final VPSD by binning into 30 logarithmically equidistant frequency bins. As for the simulations, we calculate velocities by cross-correlating the time-averaged line profile with each epoch of the time series and derive the VPSD from this.

Following the methodology within \cite{lefebvre2008variations}, we model the granulation component of each of the VPSDs using a function of the form:

\begin{equation}
    \text{$VPSD_g$}(\nu) = \frac{A_{g}}{1 + (\tau \nu)^\alpha}, \tag{7}
\end{equation}

\noindent where \( A_{g} \) is the amplitude, \( \tau \) is the characteristic timescale, and \( \alpha \) is the power-law slope. Following \cite{al2023stellar}, we fix the power-law slope \( \alpha \) = 2.

We describe the p-mode component of the VPSDs with a Lorentzian function:

\begin{equation}
    \text{$VPSD_p$}(\nu) = \frac{A_{p} \Gamma^2}{(\nu - \nu_0)^2 + \Gamma^2}, \tag{6}
\end{equation}
where \( A_{p} \) is the amplitude, \( \Gamma \) is the full-width at half-maximum (FWHM), and \( \nu_0 \) is the central frequency. 

For the simulated data, we perform a simultaneous joint fit to the original and reconstructed datasets using MCMC sampling. The original data is modeled with \( VPSD_g + VPSD_p\), while the reconstructed data is fitted with \(VPSD_g\), ensuring shared parameters for \(VPSD_g\) between the datasets. We model the observed VPSD from LARS separately with \( VPSD_g + VPSD_p + C\), where \(C\) is a constant relating to the photon noise. 

The log-likelihood for each fit is defined as: 

\begin{align}
    \ln \mathcal{L} = -\frac{1}{2} \sum_i \left[ \frac{(\log_{10}(y_i) - \text{model}_i)^2}{\sigma_i^2} + \ln(2\pi \sigma_i^2) \right],
\end{align}
where:
\begin{itemize}
    \item \(y_i\) is the observed data for dataset \(i\),
    \item \(\text{model}_i\) is the corresponding model prediction,
    \item \(\sigma_i\) represents the uncertainty in the data for dataset \(i\).
\end{itemize}

\noindent The \(\sigma_1\) and \(\sigma_2\) parameters correspond to the uncertainties (standard deviations) of the two datasets and are treated as free parameters in the MCMC sampling. The details of the MCMC analysis, including the priors and best-fit results can be found in Appendix A.

The wavelengths of the dominant p-modes in our simulations are defined by the vertical extent of the HD box. In addition, there are artifical modes excited during the relaxation process. We therefore do not expect the p-mode component of our simulated dataset to match observations (see \citealt{Zhou2021}). However, if our reconstruction method has successfully eliminated the artificial p-mode contribution, we expect the granulation component of the two datasets to be consistent. Comparing absolute amplitudes is not meaningful here, instead we care about the contribution of granulation towards the total amplitude. We therefore normalise all data and fitted components by the relevant \(A_{p} + A_{g}+C\) (in the case of simulation, \(C\) = 0). The results are shown in Figure \ref{fig:power}, with the bottom plot showing the granulation component fitted to our reconstructed data along with the component extracted from the observed data. We find a strong match to observations, confirming that our parameterisation and reconstruction method effectively isolates the granulation signature from our simulated data.

\section{Center to limb effects of granulation}
\label{sec:limb}

\begin{figure*}
    \centering
    \includegraphics[width = \textwidth] {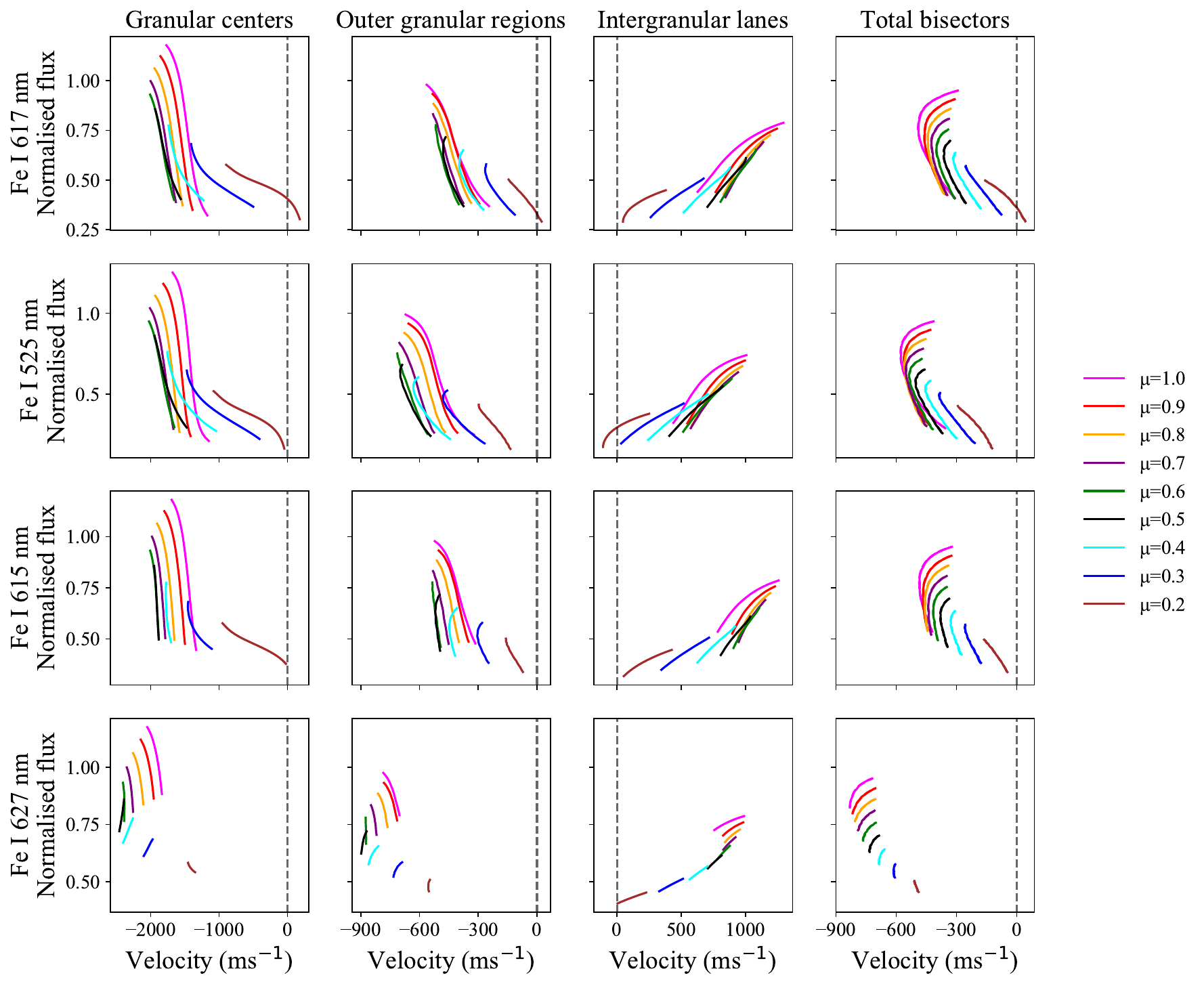}
    \caption{Bisectors of the average component profiles for each line and limb angle. Flux has been normalised relative to the continuum intensity of the total average profile at disk center (hence flux values greater than 1 can be seen in components brighter than the average). Velocity here is the Doppler velocity. The rightmost plots show the total bisectors found by combining the components multiplied by the relevant filling factors. The gray vertical dashed line indicates a velocity of 0~m/s.}
    \label{fig:comp_bis_limbs}
\end{figure*}

\begin{figure}
    \centering
    \includegraphics[width=\columnwidth]{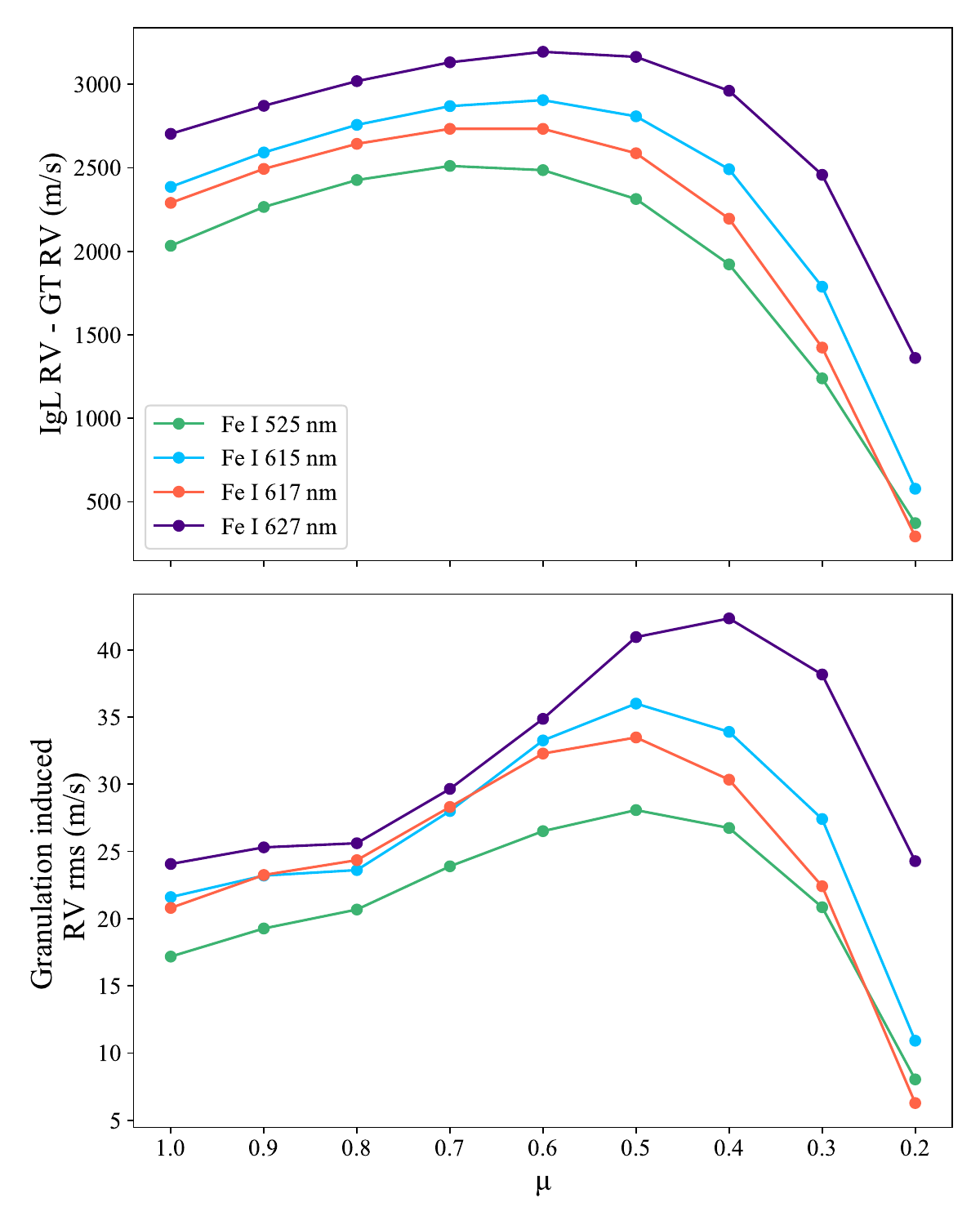}
    \caption{The top panel shows the difference in RV between intergranular lanes and granular tops for each line at different limb angles. The RVs are calculated by cross-correlating the relevant average component profile with the average total profile at each $\mu$. Note that \ion{Fe}{I}  627~nm is the weakest line and \ion{Fe}{I}  525~nm is the strongest. The bottom panel shows the changes in granulation induced RV rms. This is calculated using RVs from the reconstructed time series and indicates the level of variation between lines in granulation induced convective blueshift.}
    \label{fig:rvs_diff}
\end{figure}

\begin{figure*}
    \centering
    \includegraphics[width = \textwidth] {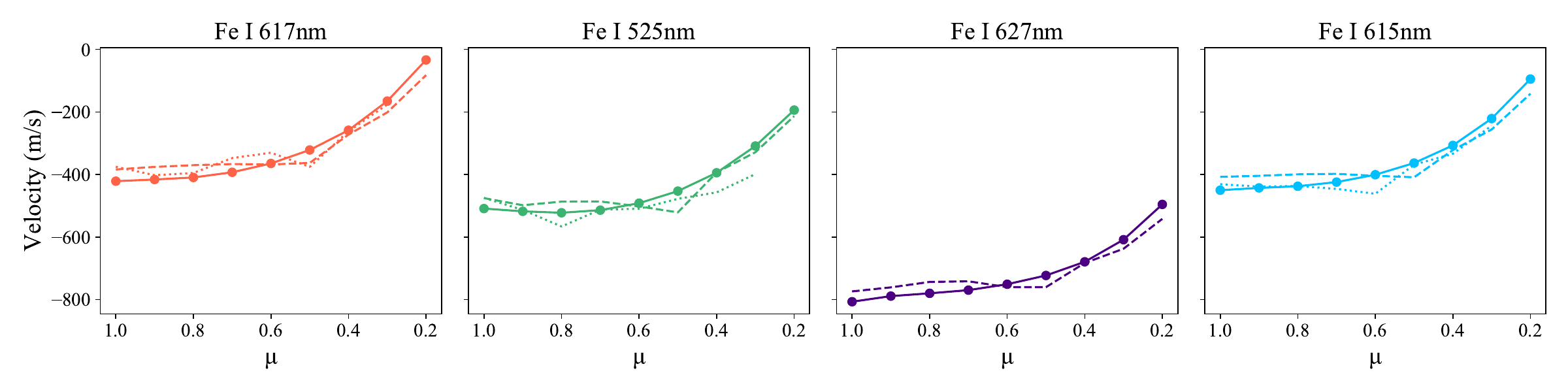}
    \caption{Variation of relative convective blueshift with limb angle, with comparisons to observations. Velocities here are the mean bisector Doppler shifts. Solid lines shows results from the reconstructed data from this work, dotted lines show results from LARS, dashed lines show results from IAG. Both observational results have been shifted to match the mean velocity across limb angles to our data. The focus of this comparison is center-to-limb variations rather than absolute velocity values.}
    \label{fig:vel_com}
\end{figure*}

\begin{figure*}
    \centering
    \includegraphics[width = \textwidth] {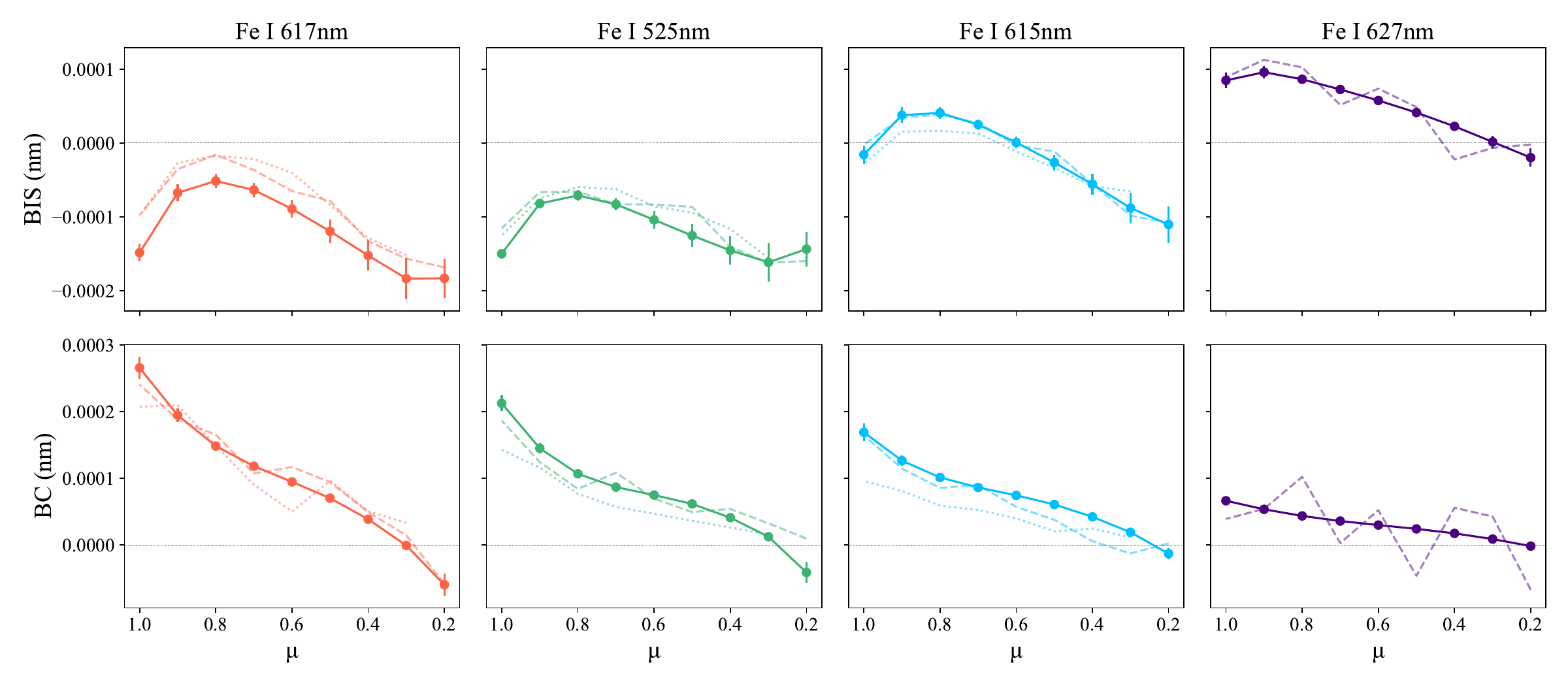}
    \caption{Bisector asymmetry measures plotted against limb angle for each line, with comparisons to observations. Bisector invserse slope (BIS) refers to the difference in average wavelength between the top (55-90\%) and bottom (10-40\%) of the bisector. Bisector curvature (BC) is the sum of the average wavelength in the top (80-90\%) and bottom (10-20\%) with the average wavelength in the middle (40-60\%) subtracted. Errorbars indicate the standard deviation across snapshots. Solid lines show results from the reconstructed data from this work. Dashed lines show results from IAG observations. Dotted lines show results from LARS observations.}
    \label{fig:asymmetry}
\end{figure*}

In this section, we use our reconstruction results to analyse granulation effects across the stellar limb for our four chosen lines. As briefly discussed in Section \ref{sec:diff_inc}, the act of inclining our line of sight changes what areas are visible due to the corrugated nature of the stellar surface. It also changes the atmospheric layer visible to us; higher inclinations probe higher altitudes, which are on average cooler and contain different velocity fields. A key consequence of this is a change in contrast between granules and lanes, which affects the bisector asymmetry and RV root mean square (rms) as a function of limb angle. 

Figure \ref{fig:comp_bis_limbs} illustrates the evolution of bisector shape and Doppler velocity for each component at different limb angles. The components displayed here are from the spatially and temporally averaged profiles solved for in Section \ref{sec:para}. The rightmost plots in this figure show the total bisectors calculated from the combination of these components. The strong characteristic C-shape at limb angles close to disk center are a direct result of the contrast in brightness and velocity of the granular and lane components. The C-shape diminishes towards the limb as the contrast decreases. Note that the bisector shapes of \ion{Fe}{I}  627~nm, the weakest line, resemble the upper portions of the stronger line bisectors, as is presented in \cite{Gray_2005}. A similar behavior can be seen in the '\ion{Fe}{I}, G1V' plot in Figure 6 in \cite{Dravins2021}. However, it is shown in \cite{Palumbo2021} that this effect is not universal across all line profiles.

The top panel in Figure \ref{fig:rvs_diff} shows the velocity contrast between Igls and GTs as a function of $\mu$. The velocities are derived by cross-correlating the component profiles with the time-averaged original profile at each $\mu$. Note that the contrast is larger for weaker lines. You can observe the reason for this in Figure \ref{fig:comp_bis_limbs}, as it is the component cores that are most similar in velocity, and weaker lines only retain the upper portion of the stronger bisectors. This contrast is a strong indicator of granulation induced RV rms, as can be seen in the bottom plot of Figure \ref{fig:rvs_diff}. These rms values are calculated using the RVs from the reconstructed time series and indicate the variability we see in the convective blueshift (CB) caused by granulation. This decrease in granular contrast for stronger lines is in agreement with \cite{Dravins2021}, who points out that in solar-like G-type stars, the amplitudes of granule brightness and velocities are somewhat obscured beneath line-forming layers. The contrast therefore decreases with height, and so stronger lines, which form higher up in the photosphere, exhibit less granular contrast. It is for this same reason that the contrast in all lines decreases so rapidly towards the limb, where higher atmospheric layers are probed. 

The actual values of CB at each $\mu$ are shown in Figure \ref{fig:vel_com}. These velocities are calculated by taking the mean bisector Doppler velocity of the time-averaged profile, where Doppler velocity is calculated relative to the rest wavelength of the line. We perform the same calculations on time-averaged disk-resolved profiles from LARS and IAG. Instead of trying to match the absolute values between our models and the observations, we focus on comparing the center-to-limb variation in CB. We therefore shift the observational results for each line to match the mean CB across $\mu$ values of simulations. The observational results are also plotted in Figure \ref{fig:vel_com}, with dashed lines being IAG and dotted LARS. Evidently, our results retain the observed relationship between $\mu$ and CB, with velocity increasing at an accelerating rate when moving towards the limb. 

Finally, in Figure \ref{fig:asymmetry}, we plot the bisector wavelength span (BIS) and bisector curvature (BC) \citep{Povich2001} against limb angle. These parameters are a way of quantifying the asymmetry of the line. They are defined as follows:

\begin{itemize}
    \item \textbf{BIS:} The average bisector wavelength within the bottom portion of the bisector subtracted from the average wavelength in the top portion. The regions are defined as 10\%-40\% and 55\%-90\% of the line depth.
    \item \textbf{BC:} The bisector is split into three regions and the average wavelength is calculated for each (top, middle, bottom). BC is calculated with top + bottom - 2$\times$ middle. Regions are defined as 80\%-90\%, 40\%-60\% and 10\%-20\%.
\end{itemize}

We calculate these measures for each profile in our reconstructed time series for each limb angle. The points plotted in Figure \ref{fig:asymmetry} are the mean values over time, and the errorbars are the standard deviations. The dashed lines show results calculated as defined above using profiles from IAG, and the dotted lines using profiles from LARS. In this plot, the observed values have not been shifted. Overall, our results align well with observations.

\section{Summary and conclusions}
\label{sec:sum}

\begin{itemize}
    \item For four selected \ion{Fe}{I} lines (\ion{Fe}{I} 525~nm, \ion{Fe}{I} 615~nm, \ion{Fe}{I} 617~nm, \ion{Fe}{I} 627~nm), we synthesize a time series at nine limb angles across the surface of a solar-type star. We use 3D HD cubes from \texttt{MURaM} and radiative transport calculations from \texttt{MPS-ATLAS}. Our resulting time series for each line consists of 129 snapshots with a cadence of 30~seconds, covering just over 1~hour. 
    
    \item The time series we generate contains the effects of p-modes and granulation. The goal of this work is to disentangle the two phenomena and isolate the impact of granulation.

    \item We parameterise the profiles based on components derived from spatial granulation variation. In this way, we can mitigate p-modes by averaging components over time, whilst maintaining the signatures of granulation through component filling factors. 
    
    \item We choose to separate the stellar surface into three components: granular tops, outer granular regions and intergranular lanes. 

    \item Our new method bases classification decisions on three spectral characteristics: continuum intensity, radial velocity and line depth. This classification is line and inclination dependent, and is not influenced by the phase of pressure modes

    \item We reconstruct our time series by averaging our three components over time and multiplying the results by the relevant filling factors at each snapshots and combining. The result is a time series that only encapsulates the impact on granulation. 

    \item We validate this by computing a Lorentzian fit to the reconstructed time series VPSD for disk center \ion{Fe}{I}  617~nm. We compare this to the granulation component extracted from LARS observations of the same line at the solar disk center. We find an excellent match, strongly suggesting that our method has been successful in isolating granulation. 

    \item We use our isolated granulation time series to analyse center to limb variation of granulation effects. We show that velocity contrast between granules and lanes slowly increases when moving from disk center to $\mu \approx 0.7$, before flattening off and then sharply decreasing at $\mu \approx 0.5$. The contrast gets stronger the weaker the line, resulting in higher granulation induced RV rms for weak lines. 

    \item We show that the relationship between convective blueshift and limb angle is generally consistent with observations. 

    \item We quantify bisector asymmetry with two measures and show line by line differences as well as center to limb variations. We find strong matches with observations. 
    
\end{itemize}

\section{Future work}

In this paper, we have demonstrated that we are able to use parameterisation to effectively isolate granulation signatures from 3D HD models. Not only does this allow us to study uncontaminated granulation effects on spectral lines, it also provides a computationally efficient way of generating spectra. Using our average component profiles and the distributions of filling factors, we can generate an unlimited number of profiles quickly. See \cite{cegla2019} for an example of this. This paves the way towards tiling a full stellar disk and producing disk-integrated spectra for unlimited instances in time. Using our isolated granulation disk-integrated spectra, we can study relationships between bisector asymmetry and granulation-induced convective blueshift, as in \cite{Palumbo2021}. These correlations could play a vital part in mitigating granulation induced noise in RV measurements, allowing us to detect smaller, longer period planets. 

A key benefit of conducting this study using simulations is the ability to alter the conditions within our 3D cubes. We intend to extend this work to include various magnetic field strengths and quantify the impacts of magnetic field on granulation effects. 

\section{Acknowledgments}

This work has made use of the VALD database, operated at Uppsala University, the Institute of Astronomy RAS in Moscow, and the University of Vienna.

The Vacuum Tower Telescope at the Observatorio del Teide on Tenerife is operated by the Leibniz Institute for Solar Physics (KIS) Freiburg which is a public-law foundation of the state of Baden-W\"urttemberg and member of the Leibniz-Gemeinschaft. The installation and characterization of LARS at the VTT were funded by Leibniz-Gemeinschaft from 2011 to 2014 through the ``Pakt f\"ur Forschung und Innovation." The initial characterization of the laser frequency comb at the VTT was based on an agreement between Max-Planck-Institute for Quantum Optics (Garching) and KIS. The scientific exploitation of LARS was supported by Deutsche Forschungsgemeinschaft under grant Schm-1168/10 between 2016 and 2018.

This work acknowledges funding from a UKRI Future Leader Fellowship (grant numbers MR/S035214/1 and MR/Y011759/1). GF acknowledges a Warwick prize scholarship (PhD) made possible thanks to a generous philanthropic donation. MLP was supported by the Flatiron Research Fellowship at the Flatiron Institute, a division of the Simons Foundation. CAW would like to acknowledge support from the UK Science and Technology Facilities Council (STFC, grant number ST/X00094X/1). AIS and VW acknowledge support from the European Research Council (ERC) under the European Union’s Horizon 2020 research and innovation program (grant no. 101118581). 

We would like to thank the anonymous referee for helpful comments.

\section{Data Availability}

Datasets generated during this study are available upon reasonable request.



\bibliographystyle{mnras}
\bibliography{bib} 

\begin{thebibliography}{}
\makeatletter
\relax
\def\mn@urlcharsother{\let\do\@makeother \do\$\do\&\do\#\do\^\do\_\do\%\do\~}
\def\mn@doi{\begingroup\mn@urlcharsother \@ifnextchar [ {\mn@doi@} {\mn@doi@[]}}
\def\mn@doi@[#1]#2{\def\@tempa{#1}\ifx\@tempa\@empty \href {http://dx.doi.org/#2} {doi:#2}\else \href {http://dx.doi.org/#2} {#1}\fi \endgroup}
\def\mn@eprint#1#2{\mn@eprint@#1:#2::\@nil}
\def\mn@eprint@arXiv#1{\href {http://arxiv.org/abs/#1} {{\tt arXiv:#1}}}
\def\mn@eprint@dblp#1{\href {http://dblp.uni-trier.de/rec/bibtex/#1.xml} {dblp:#1}}
\def\mn@eprint@#1:#2:#3:#4\@nil{\def\@tempa {#1}\def\@tempb {#2}\def\@tempc {#3}\ifx \@tempc \@empty \let \@tempc \@tempb \let \@tempb \@tempa \fi \ifx \@tempb \@empty \def\@tempb {arXiv}\fi \@ifundefined {mn@eprint@\@tempb}{\@tempb:\@tempc}{\expandafter \expandafter \csname mn@eprint@\@tempb\endcsname \expandafter{\@tempc}}}

\bibitem[\protect\citeauthoryear{{Al Moulla}, {Dumusque}, {Cretignier}, {Zhao}  \& {Valenti}}{{Al Moulla} et~al.}{2022}]{moulla}
{Al Moulla} K.,  {Dumusque} X.,  {Cretignier} M.,  {Zhao} Y.,   {Valenti} J.~A.,  2022, \mn@doi [\aap] {10.1051/0004-6361/202243276}, \href {https://ui.adsabs.harvard.edu/abs/2022A&A...664A..34A} {664, A34}

\bibitem[\protect\citeauthoryear{{Al Moulla}, {Dumusque}, {Figueira}, {Lo Curto}, {Santos}  \& {Wildi}}{{Al Moulla} et~al.}{2023a}]{al2023stellar}
{Al Moulla} K.,  {Dumusque} X.,  {Figueira} P.,  {Lo Curto} G.,  {Santos} N.~C.,   {Wildi} F.,  2023a, \mn@doi [\aap] {10.1051/0004-6361/202244663}, \href {https://ui.adsabs.harvard.edu/abs/2023A&A...669A..39A} {669, A39}

\bibitem[\protect\citeauthoryear{{Al Moulla}, {Dumusque}, {Figueira}, {Lo Curto}, {Santos}  \& {Wildi}}{{Al Moulla} et~al.}{2023b}]{Moulla_power}
{Al Moulla} K.,  {Dumusque} X.,  {Figueira} P.,  {Lo Curto} G.,  {Santos} N.~C.,   {Wildi} F.,  2023b, \mn@doi [\aap] {10.1051/0004-6361/202244663}, \href {https://ui.adsabs.harvard.edu/abs/2023A&A...669A..39A} {669, A39}

\bibitem[\protect\citeauthoryear{{Allen} et~al.,}{{Allen} et~al.}{2018}]{NEID}
{Allen} L.~E.,  et~al., 2018, in American Astronomical Society Meeting Abstracts \#231. p. 246.08

\bibitem[\protect\citeauthoryear{{Asplund}, {Grevesse}, {Sauval}  \& {Scott}}{{Asplund} et~al.}{2009}]{Asplund_2009}
{Asplund} M.,  {Grevesse} N.,  {Sauval} A.~J.,   {Scott} P.,  2009, \mn@doi [\araa] {10.1146/annurev.astro.46.060407.145222}, \href {https://ui.adsabs.harvard.edu/abs/2009ARA&A..47..481A} {47, 481}

\bibitem[\protect\citeauthoryear{{Bard} \& {Kock}}{{Bard} \& {Kock}}{1994}]{BK}
{Bard} A.,  {Kock} M.,  1994, \aap, 282, 1014

\bibitem[\protect\citeauthoryear{{Barklem}, {Piskunov}  \& {O'Mara}}{{Barklem} et~al.}{2000}]{BPM}
{Barklem} P.~S.,  {Piskunov} N.,   {O'Mara} B.~J.,  2000, \mn@doi [\aap] {10.1051/aas:2000167}, 142, 467

\bibitem[\protect\citeauthoryear{{Cegla}}{{Cegla}}{2019}]{ceglareview}
{Cegla} H.~M.,  2019, \mn@doi [Geosciences] {10.3390/geosciences9030114}, \href {https://ui.adsabs.harvard.edu/abs/2019Geosc...9..114C} {9, 114}

\bibitem[\protect\citeauthoryear{{Cegla}, {Shelyag}, {Watson}  \& {Mathioudakis}}{{Cegla} et~al.}{2013}]{cegla2013}
{Cegla} H.~M.,  {Shelyag} S.,  {Watson} C.~A.,   {Mathioudakis} M.,  2013, \mn@doi [\apj] {10.1088/0004-637X/763/2/95}, \href {https://ui.adsabs.harvard.edu/abs/2013ApJ...763...95C} {763, 95}

\bibitem[\protect\citeauthoryear{{Cegla} et~al.,}{{Cegla} et~al.}{2018}]{cegla2018}
{Cegla} H.~M.,  et~al., 2018, \mn@doi [\apj] {10.3847/1538-4357/aaddfc}, \href {https://ui.adsabs.harvard.edu/abs/2018ApJ...866...55C} {866, 55}

\bibitem[\protect\citeauthoryear{{Cegla}, {Watson}, {Shelyag}, {Mathioudakis}  \& {Moutari}}{{Cegla} et~al.}{2019}]{cegla2019}
{Cegla} H.~M.,  {Watson} C.~A.,  {Shelyag} S.,  {Mathioudakis} M.,   {Moutari} S.,  2019, \mn@doi [\apj] {10.3847/1538-4357/ab16d3}, \href {https://ui.adsabs.harvard.edu/abs/2019ApJ...879...55C} {879, 55}

\bibitem[\protect\citeauthoryear{{Chiavassa} \& {Brogi}}{{Chiavassa} \& {Brogi}}{2019}]{Chiavassa2019}
{Chiavassa} A.,  {Brogi} M.,  2019, \mn@doi [\aap] {10.1051/0004-6361/201936566}, \href {https://ui.adsabs.harvard.edu/abs/2019A&A...631A.100C} {631, A100}

\bibitem[\protect\citeauthoryear{{Cretignier}, {Dumusque}, {Allart}, {Pepe}  \& {Lovis}}{{Cretignier} et~al.}{2020}]{Cretignier2020}
{Cretignier} M.,  {Dumusque} X.,  {Allart} R.,  {Pepe} F.,   {Lovis} C.,  2020, \mn@doi [\aap] {10.1051/0004-6361/201936548}, \href {https://ui.adsabs.harvard.edu/abs/2020A&A...633A..76C} {633, A76}

\bibitem[\protect\citeauthoryear{{Dravins}}{{Dravins}}{1990}]{Dravins1990}
{Dravins} D.,  1990, \aap, \href {https://ui.adsabs.harvard.edu/abs/1990A&A...228..218D} {228, 218}

\bibitem[\protect\citeauthoryear{{Dravins}}{{Dravins}}{2008}]{Dravins2008}
{Dravins} D.,  2008, \mn@doi [\aap] {10.1051/0004-6361:200810481}, \href {https://ui.adsabs.harvard.edu/abs/2008A&A...492..199D} {492, 199}

\bibitem[\protect\citeauthoryear{{Dravins} \& {Ludwig}}{{Dravins} \& {Ludwig}}{2023}]{Dravins2023}
{Dravins} D.,  {Ludwig} H.-G.,  2023, \mn@doi [\aap] {10.1051/0004-6361/202347142}, \href {https://ui.adsabs.harvard.edu/abs/2023A&A...679A...3D} {679, A3}

\bibitem[\protect\citeauthoryear{{Dravins}, {Lindegren}  \& {Nordlund}}{{Dravins} et~al.}{1981}]{Dravins1981}
{Dravins} D.,  {Lindegren} L.,   {Nordlund} A.,  1981, \aap, \href {https://ui.adsabs.harvard.edu/abs/1981A&A....96..345D} {96, 345}

\bibitem[\protect\citeauthoryear{{Dravins}, {Ludwig}  \& {Freytag}}{{Dravins} et~al.}{2021}]{Dravins2021}
{Dravins} D.,  {Ludwig} H.-G.,   {Freytag} B.,  2021, \mn@doi [\aap] {10.1051/0004-6361/202039995}, \href {https://ui.adsabs.harvard.edu/abs/2021A&A...649A..16D} {649, A16}

\bibitem[\protect\citeauthoryear{Dumusque}{Dumusque}{2018}]{Dumusque_2018}
Dumusque X.,  2018, \mn@doi [{\aap}] {10.1051/0004-6361/201833795}, 620, A47

\bibitem[\protect\citeauthoryear{{Dumusque}, {Udry}, {Lovis}, {Santos}  \& {Monteiro}}{{Dumusque} et~al.}{2011}]{Dumusque}
{Dumusque} X.,  {Udry} S.,  {Lovis} C.,  {Santos} N.~C.,   {Monteiro} M.~J.~P.~F.~G.,  2011, \mn@doi [\aap] {10.1051/0004-6361/201014097}, \href {https://ui.adsabs.harvard.edu/abs/2011A&A...525A.140D} {525, A140}

\bibitem[\protect\citeauthoryear{{Ellwarth}, {Sch{\"a}fer}, {Reiners}  \& {Zechmeister}}{{Ellwarth} et~al.}{2023}]{IAG}
{Ellwarth} M.,  {Sch{\"a}fer} S.,  {Reiners} A.,   {Zechmeister} M.,  2023, \mn@doi [\aap] {10.1051/0004-6361/202245612}, \href {https://ui.adsabs.harvard.edu/abs/2023A&A...673A..19E} {673, A19}

\bibitem[\protect\citeauthoryear{Foreman-Mackey}{Foreman-Mackey}{2014}]{foreman2014blog}
Foreman-Mackey D.,  2014, Blog Post: Mixture Models, doi: 10.5281/zenodo. 15856

\bibitem[\protect\citeauthoryear{{Fuhr}, {Martin}  \& {Wiese}}{{Fuhr} et~al.}{1988}]{FMW}
{Fuhr} J.~R.,  {Martin} G.~A.,   {Wiese} W.~L.,  1988, Journal of Physical and Chemical Reference Data, Volume 17, Suppl.~4.~New York: American Institute of Physics (AIP) and American Chemical Society, 1988, \href {http://cdsads.u-strasbg.fr/abs/1988JPCRD..17S....F} {17}

\bibitem[\protect\citeauthoryear{Gray}{Gray}{2005}]{Gray_2005}
Gray D.~F.,  2005, Velocity fields in stellar photospheres.
Cambridge University Press, p. 423–457

\bibitem[\protect\citeauthoryear{{Irwin}}{{Irwin}}{2012}]{Irwin_freeeos_2012}
{Irwin} A.~W.,  2012, {FreeEOS: Equation of State for stellar interiors calculations} (\mn@eprint {ascl} {1211.002})

\bibitem[\protect\citeauthoryear{Jurgenson, Fischer, McCracken, Sawyer, Szymkowiak, Davis, Muller  \& Santoro}{Jurgenson et~al.}{2016}]{expres}
Jurgenson C.,  Fischer D.,  McCracken T.,  Sawyer D.,  Szymkowiak A.,  Davis A.,  Muller G.,   Santoro F.,  2016, in Ground-based and airborne instrumentation for astronomy vi. pp 2051--2070

\bibitem[\protect\citeauthoryear{{Kallinger} et~al.,}{{Kallinger} et~al.}{2014}]{Kallinger2014}
{Kallinger} T.,  et~al., 2014, \mn@doi [\aap] {10.1051/0004-6361/201424313}, \href {https://ui.adsabs.harvard.edu/abs/2014A&A...570A..41K} {570, A41}

\bibitem[\protect\citeauthoryear{{Kurucz}}{{Kurucz}}{2014}]{K14}
{Kurucz} R.~L.,  2014, Robert L. Kurucz on-line database of observed and predicted atomic transitions

\bibitem[\protect\citeauthoryear{{Lafarga} et~al.,}{{Lafarga} et~al.}{2023}]{Lafarga_2023}
{Lafarga} M.,  et~al., 2023, \mn@doi [\aap] {10.1051/0004-6361/202245602}, \href {https://ui.adsabs.harvard.edu/abs/2023A&A...674A..61L} {674, A61}

\bibitem[\protect\citeauthoryear{Lefebvre, Garc{\'\i}a, Jim{\'e}nez-Reyes, Turck-Chi{\`e}ze  \& Mathur}{Lefebvre et~al.}{2008}]{lefebvre2008variations}
Lefebvre S.,  Garc{\'\i}a R.,  Jim{\'e}nez-Reyes S.,  Turck-Chi{\`e}ze S.,   Mathur S.,  2008, \aap, 490, 1143

\bibitem[\protect\citeauthoryear{{L{\"o}hner-B{\"o}ttcher}, {Schmidt}, {Doerr}, {Kentischer}, {Steinmetz}, {Probst}  \& {Holzwarth}}{{L{\"o}hner-B{\"o}ttcher} et~al.}{2017}]{Lohner-Bottcher2017}
{L{\"o}hner-B{\"o}ttcher} J.,  {Schmidt} W.,  {Doerr} H.~P.,  {Kentischer} T.,  {Steinmetz} T.,  {Probst} R.~A.,   {Holzwarth} R.,  2017, \mn@doi [\aap] {10.1051/0004-6361/201731164}, \href {https://ui.adsabs.harvard.edu/abs/2017A&A...607A..12L} {607, A12}

\bibitem[\protect\citeauthoryear{{L{\"o}hner-B{\"o}ttcher}, {Schmidt}, {Stief}, {Steinmetz}  \& {Holzwarth}}{{L{\"o}hner-B{\"o}ttcher} et~al.}{2018}]{LARS}
{L{\"o}hner-B{\"o}ttcher} J.,  {Schmidt} W.,  {Stief} F.,  {Steinmetz} T.,   {Holzwarth} R.,  2018, \mn@doi [\aap] {10.1051/0004-6361/201732107}, \href {https://ui.adsabs.harvard.edu/abs/2018A&A...611A...4L} {611, A4}

\bibitem[\protect\citeauthoryear{L{\"o}hner-B{\"o}ttcher, Schmidt, Schlichenmaier, Steinmetz  \& Holzwarth}{L{\"o}hner-B{\"o}ttcher et~al.}{2019}]{lohner2019}
L{\"o}hner-B{\"o}ttcher J.,  Schmidt W.,  Schlichenmaier R.,  Steinmetz T.,   Holzwarth R.,  2019, \aap, 624, A57

\bibitem[\protect\citeauthoryear{{Magic}, {Collet}, {Asplund}, {Trampedach}, {Hayek}, {Chiavassa}, {Stein}  \& {Nordlund}}{{Magic} et~al.}{2013}]{Magic_2013A&A}
{Magic} Z.,  {Collet} R.,  {Asplund} M.,  {Trampedach} R.,  {Hayek} W.,  {Chiavassa} A.,  {Stein} R.~F.,   {Nordlund} {\AA}.,  2013, \mn@doi [\aap] {10.1051/0004-6361/201321274}, 557, A26

\bibitem[\protect\citeauthoryear{{Meunier}, {Lagrange}, {Borgniet}  \& {Rieutord}}{{Meunier} et~al.}{2015}]{Meunier2015}
{Meunier} N.,  {Lagrange} A.~M.,  {Borgniet} S.,   {Rieutord} M.,  2015, \mn@doi [\aap] {10.1051/0004-6361/201525721}, \href {https://ui.adsabs.harvard.edu/abs/2015A&A...583A.118M} {583, A118}

\bibitem[\protect\citeauthoryear{{Michel}, {Samadi}, {Baudin}, {Barban}, {Appourchaux}  \& {Auvergne}}{{Michel} et~al.}{2009}]{Michel2009}
{Michel} E.,  {Samadi} R.,  {Baudin} F.,  {Barban} C.,  {Appourchaux} T.,   {Auvergne} M.,  2009, \mn@doi [\aap] {10.1051/0004-6361:200810353}, \href {https://ui.adsabs.harvard.edu/abs/2009A&A...495..979M} {495, 979}

\bibitem[\protect\citeauthoryear{{Palumbo}, {Ford}, {Wright}, {Mahadevan}, {Wise}  \& {L{\"o}hner-B{\"o}ttcher}}{{Palumbo} et~al.}{2022}]{grass1}
{Palumbo} III M.~L.,  {Ford} E.~B.,  {Wright} J.~T.,  {Mahadevan} S.,  {Wise} A.~W.,   {L{\"o}hner-B{\"o}ttcher} J.,  2022, \mn@doi [\aj] {10.3847/1538-3881/ac32c2}, \href {https://ui.adsabs.harvard.edu/abs/2022AJ....163...11P} {163, 11}

\bibitem[\protect\citeauthoryear{{Palumbo}, {Ford}, {Gonzalez}, {Wright}, {Al Moulla}  \& {Schlichenmaier}}{{Palumbo} et~al.}{2024a}]{grass2}
{Palumbo} M.~L.,  {Ford} E.~B.,  {Gonzalez} E.~B.,  {Wright} J.~T.,  {Al Moulla} K.,   {Schlichenmaier} R.,  2024a, \mn@doi [\aj] {10.3847/1538-3881/ad4c6d}, \href {https://ui.adsabs.harvard.edu/abs/2024AJ....168...46P} {168, 46}

\bibitem[\protect\citeauthoryear{{Palumbo}, {Ford}, {Gonzalez}, {Wright}, {Al Moulla}  \& {Schlichenmaier}}{{Palumbo} et~al.}{2024b}]{Palumbo2021}
{Palumbo} M.~L.,  {Ford} E.~B.,  {Gonzalez} E.~B.,  {Wright} J.~T.,  {Al Moulla} K.,   {Schlichenmaier} R.,  2024b, \mn@doi [\aj] {10.3847/1538-3881/ad4c6d}, \href {https://ui.adsabs.harvard.edu/abs/2024AJ....168...46P} {168, 46}

\bibitem[\protect\citeauthoryear{Pedregosa et~al.,}{Pedregosa et~al.}{2011}]{scikit-learn}
Pedregosa F.,  et~al., 2011, Journal of Machine Learning Research, 12, 2825

\bibitem[\protect\citeauthoryear{{Pepe} et~al.,}{{Pepe} et~al.}{2013}]{espresso}
{Pepe} F.,  et~al., 2013, The Messenger, \href {https://ui.adsabs.harvard.edu/abs/2013Msngr.153....6P} {153, 6}

\bibitem[\protect\citeauthoryear{{Povich}, {Giampapa}, {Valenti}, {Tilleman}, {Barden}, {Deming}, {Livingston}  \& {Pilachowski}}{{Povich} et~al.}{2001}]{Povich2001}
{Povich} M.~S.,  {Giampapa} M.~S.,  {Valenti} J.~A.,  {Tilleman} T.,  {Barden} S.,  {Deming} D.,  {Livingston} W.~C.,   {Pilachowski} C.,  2001, \mn@doi [\aj] {10.1086/318745}, \href {https://ui.adsabs.harvard.edu/abs/2001AJ....121.1136P} {121, 1136}

\bibitem[\protect\citeauthoryear{Rackham et~al.,}{Rackham et~al.}{2023}]{NASA}
Rackham B.~V.,  et~al., 2023, The Effect of Stellar Contamination on Low-resolution Transmission Spectroscopy: Needs Identified by NASA's Exoplanet Exploration Program Study Analysis Group 21 (\mn@eprint {arXiv} {2201.09905}), \url {https://arxiv.org/abs/2201.09905}

\bibitem[\protect\citeauthoryear{{Rempel}}{{Rempel}}{2014}]{rempel2014}
{Rempel} M.,  2014, \mn@doi [\apj] {10.1088/0004-637X/789/2/132}, \href {https://ui.adsabs.harvard.edu/abs/2014ApJ...789..132R} {789, 132}

\bibitem[\protect\citeauthoryear{{Schrijver} \& {Zwaan}}{{Schrijver} \& {Zwaan}}{2000}]{book}
{Schrijver} C.~J.,  {Zwaan} C.,  2000, {Solar and stellar magnetic activity. Cambridge Astrophysics Series.}

\bibitem[\protect\citeauthoryear{{Socas-Navarro}}{{Socas-Navarro}}{2015}]{nicole1}
{Socas-Navarro} H.,  2015, {NICOLE: NLTE Stokes Synthesis/Inversion Code}, Astrophysics Source Code Library, record ascl:1508.002

\bibitem[\protect\citeauthoryear{{Socas-Navarro}, {de la Cruz Rodr{\'\i}guez}, {Asensio Ramos}, {Trujillo Bueno}  \& {Ruiz Cobo}}{{Socas-Navarro} et~al.}{2015}]{nicole2}
{Socas-Navarro} H.,  {de la Cruz Rodr{\'\i}guez} J.,  {Asensio Ramos} A.,  {Trujillo Bueno} J.,   {Ruiz Cobo} B.,  2015, \mn@doi [\aap] {10.1051/0004-6361/201424860}, \href {https://ui.adsabs.harvard.edu/abs/2015A&A...577A...7S} {577, A7}

\bibitem[\protect\citeauthoryear{{Takeda} \& {UeNo}}{{Takeda} \& {UeNo}}{2017}]{Yoichi}
{Takeda} Y.,  {UeNo} S.,  2017, \mn@doi [\pasj] {10.1093/pasj/psx022}, \href {https://ui.adsabs.harvard.edu/abs/2017PASJ...69...46T} {69, 46}

\bibitem[\protect\citeauthoryear{{Uitenbroek} \& {Criscuoli}}{{Uitenbroek} \& {Criscuoli}}{2011}]{Han}
{Uitenbroek} H.,  {Criscuoli} S.,  2011, \mn@doi [\apj] {10.1088/0004-637X/736/1/69}, \href {https://ui.adsabs.harvard.edu/abs/2011ApJ...736...69U} {736, 69}

\bibitem[\protect\citeauthoryear{{V{\"o}gler}, {Shelyag}, {Sch{\"u}ssler}, {Cattaneo}, {Emonet}  \& {Linde}}{{V{\"o}gler} et~al.}{2005}]{MURaM}
{V{\"o}gler} A.,  {Shelyag} S.,  {Sch{\"u}ssler} M.,  {Cattaneo} F.,  {Emonet} T.,   {Linde} T.,  2005, \mn@doi [\aap] {10.1051/0004-6361:20041507}, \href {https://ui.adsabs.harvard.edu/abs/2005A&A...429..335V} {429, 335}

\bibitem[\protect\citeauthoryear{{Witzke} et~al.,}{{Witzke} et~al.}{2021}]{mps-atlas}
{Witzke} V.,  et~al., 2021, \mn@doi [\aap] {10.1051/0004-6361/202140275}, \href {https://ui.adsabs.harvard.edu/abs/2021A&A...653A..65W} {653, A65}

\bibitem[\protect\citeauthoryear{{Witzke} et~al.,}{{Witzke} et~al.}{2024}]{witzke2024}
{Witzke} V.,  et~al., 2024, \mn@doi [\aap] {10.1051/0004-6361/202346099}, \href {https://ui.adsabs.harvard.edu/abs/2024A&A...681A..81W} {681, A81}

\bibitem[\protect\citeauthoryear{{Zhou}, {Nordlander}, {Casagrande}, {Joyce}, {Li}, {Amarsi}, {Reggiani}  \& {Asplund}}{{Zhou} et~al.}{2021}]{Zhou2021}
{Zhou} Y.,  {Nordlander} T.,  {Casagrande} L.,  {Joyce} M.,  {Li} Y.,  {Amarsi} A.~M.,  {Reggiani} H.,   {Asplund} M.,  2021, \mn@doi [\mnras] {10.1093/mnras/stab337}, \href {https://ui.adsabs.harvard.edu/abs/2021MNRAS.503...13Z} {503, 13}

\makeatother
\end{thebibliography}



\appendix 
\section{Velocity power spectra fit details}
\label{app:A1}

In the following we present details and results for the MCMC sampling discussed in Section \ref{sec:power}.

The python package \texttt{emcee} \citep{foreman2014blog} was used for MCMC analysis, employing 32 walkers and 10,000 steps per walker, with a burn-in phase of 1500 steps. Table~\ref{tab:priors} outlines the priors for the parameters, while Table~\ref{tab:best_fit_parameters} presents the best-fit results for both the simulated and observed dataset. Note that the characteristic timescale of granulation ($\tau$) will be affected by the spatial scale of observations. The values presented here for disk center calculations should not be compared to those from disk-integrated studies (e.g: \cite{Moulla_power}). The fact that our $\tau$ result from our synthetic dataset shows a strong agreement with that derived from disk center solar observations indicates that our granulation extraction method accurately captures the temporal evolution.

\begin{table*}
\centering
\begin{tabular}{|c|c|c|c|}
\hline
\textbf{Parameter} & \textbf{Description} & \textbf{Prior (simulated)} & \textbf{Prior (observed)} \\
\hline
\multicolumn{4}{|c|}{\textbf{P-mode Parameters}} \\
\hline
\(A_{p}~[(m s^{-1})^{2}~Hz^{-1}]\) & Amplitude  & \(U[5 \times 10^7, 2 \times 10^8]\) & \(U[5 \times 10^7, 2 \times 10^8]\)\\

\(\Gamma~[mHz]\) & FWHM & \(U[0, 0.3]\) & \(U[0, 0.3]\)\\

\(\nu_{0}~[mHz]\) & Central frequency & \(U[2.8, 3.5]\) & \(U[2.8, 3.5]\)\\

\hline
\multicolumn{4}{|c|}{\textbf{Granulation Parameters}} \\
\hline

\(A_{g}~[(m s^{-1})^{2}~Hz^{-1}]\) & Amplitude & \(U[2 \times 10^4, 5 \times 10^5]\) & \(U[2 \times 10^4, 5 \times 10^5]\)\\

\(\tau~[mHz^{-1}]\) & Characteristic timescale & \(U[0.1, 1.0]\) & \(U[0.1, 1.0]\)\\

\hline
\multicolumn{4}{|c|}{\textbf{Constants}} \\
\hline

\(C~[(m s^{-1})^{2}~Hz^{-1}]\) & Photon noise & N/A & \(U[1 \times 10^2, 1 \times 10^5]\)\\

\hline
\multicolumn{4}{|c|}{\textbf{Uncertainties}} \\
\hline

\(\sigma_1~[(m s^{-1})^{2}~Hz^{-1}]\) & Uncertainty in dataset 1 & \(U[10^{-3}, 10]\) & \(U[10^{-3}, 10]\)\\

\(\sigma_2~[(m s^{-1})^{2}~Hz^{-1}]\) & Uncertainty in dataset 2 & \(U[10^{-3}, 10]\) & N/A\\

\hline
\end{tabular}
\caption{Priors for MCMC parameters. \(U[\text{min}, \text{max}]\) denotes a uniform prior.}
\label{tab:priors}
\end{table*}

\begin{table*}
\centering
\begin{tabular}{|c|c|c|c|c|}
\hline
\textbf{Parameter} & \textbf{Best Fit (simulated)} & \textbf{Uncertainty (simulated)} & \textbf{Best Fit (observed)} & \textbf{Uncertainty (observed)} \\
\hline
\multicolumn{5}{|c|}{\textbf{P-mode Parameters}} \\
\hline
\( A_{p}~[(m s^{-1})^{2}~Hz^{-1}]\) & \(1.10 \times 10^8\) & \(6.51 \times 10^7 / 1.68 \times 10^8\) & \(1.01 \times 10^8\) & \(6.28 \times 10^7 / 1.63 \times 10^8\)\\

\( \Gamma~[mHz]\) & \(5.60 \times 10^{-2}\) & \(4.17 \times 10^{-2} / 7.66 \times 10^{-2}\) & \(7.58 \times 10^{-2}\) & \(5.50 \times 10^{-2} / 1.05 \times 10^{-1}\)\\

\( \nu_0~[mHz] \) & \(2.96 \) & \(2.88 / 3.05 \) & \(3.25 \) & \(3.19/ 3.31 \)\\

\hline
\multicolumn{5}{|c|}{\textbf{Granulation Parameters}} \\
\hline
\( A_{g}~[(m s^{-1})^{2}~Hz^{-1}] \) & \(1.43 \times 10^5\) & \(9.59 \times 10^4 / 2.13 \times 10^5\) & \(3.28 \times 10^5\) & \(2.12 \times 10^5 / 4.34 \times 10^5\)\\

\( \tau~[mHz^{-1}]\) & \(3.25 \times 10^{-1}\) & \(2.55 \times 10^{-1} / 4.18 \times 10^{-1}\) & \(3.60 \times 10^{-1}\) & \(2.12 \times 10^{-1} / 4.34 \times 10^{-1}\)\\

\hline
\multicolumn{5}{|c|}{\textbf{Constants}} \\
\hline

\(C~[(m s^{-1})^{2}~Hz^{-1}]\) & N/A & N/A & \(2.85 \times 10^3\) & \(1.99 \times 10^3 / 3.96 \times 10^3\)\\

\hline
\multicolumn{5}{|c|}{\textbf{Uncertainties}} \\
\hline

\( \sigma_1~[(m s^{-1})^{2}~Hz^{-1}] \) & \(5.55 \times 10^{-1}\) & \(5.08 \times 10^{-1} / 6.08 \times 10^{-1}\) & \(3.21 \times 10^{-1}\) & \(2.82 \times 10^{-1} / 3.69 \times 10^{-1}\)\\

\( \sigma_2~[(m s^{-1})^{2}~Hz^{-1}] \) & \(6.79 \times 10^{-1}\) & \(6.22 \times 10^{-1} / 7.46 \times 10^{-1}\) & N/A & N/A \\

\hline
\end{tabular}
\caption{Best fit parameters and uncertainties from MCMC sampling. The uncertainties represent the 16th and 84th percentiles of the posterior distributions.}
\label{tab:best_fit_parameters}
\end{table*}

\bsp	
\label{lastpage}
\end{document}